\documentclass[12pt]{article}
\usepackage{amsfonts}
\usepackage{float}
\usepackage{jcappub}
\usepackage{graphicx}
\graphicspath{{figures/}} 
 \usepackage{caption} 
 \usepackage{subcaption} 
 \usepackage{epstopdf}
 \usepackage{hyperref}
 \usepackage{cleveref}  
 \usepackage{enumitem}
 \usepackage{pdfpages}
 \usepackage{multicol}  
\usepackage{mathrsfs}
\usepackage{amssymb}
\usepackage{textcomp}
\usepackage{amsmath, float}
 \usepackage{times}
 \usepackage{etoolbox}
\usepackage{color}
\usepackage[normalem]{ulem}
\usepackage{enumitem}
\usepackage{mathpazo} 
\usepackage{drsn_ref}
\usepackage{multirow}
\usepackage[english]{babel}
\usepackage[nottoc]{tocbibind}

\usepackage{comment}
\includecomment{toexclude}   

\begin{document} 

\title{\boldmath Gamma Rays Bursts: A Viable Cosmological Probe? }

\author[a]{Darshan Kumar}
\author[b]{, Nisha Rani}
\author[c]{, Deepak Jain}
\author[a]{, Shobhit Mahajan}
\author[a]{, Amitabha Mukherjee}
\affiliation[a]{\small Department of Physics and Astrophysics, University of Delhi, Delhi 110007, India}
\affiliation[b]{\small Miranda House, University of Delhi, University Enclave, Delhi-110007, India}
\affiliation[c]{\small Deen Dayal Upadhyaya College, University of Delhi, Dwarka, New Delhi 110078, India}

\emailAdd{darshanbeniwal11@gmail.com}
\emailAdd{nisha.physics@mirandahouse.ac.in}
\emailAdd{djain@ddu.du.ac.in} 
\emailAdd{sm@physics.du.ac.in}
\emailAdd{am@physics.du.ac.in}

\abstract{{
In this work, our focus is on exploring the potential of current GRB measurements to provide reliable constraints on cosmological model parameters at high redshift. This work is divided into two parts. First, we calibrate the Amati relation in a model-independent way by using Hubble parameter measurements obtained from the differential ages of the galaxies. We further check if the Amati relation parameters evolve with the GRBs' redshift or not, using the data of Old Astrophysical Objects. The results indicate that GRBs do seem to evolve  with redshift.} In the second part, we test  different cosmological models with the calibrated GRB data obtained by using constant and dynamical Amati relation. Our results indicate that the present quality of GRB data  is not good enough to put tight constraints on the cosmological parameters. Hence we perform a joint analysis with the  combined data of GRBs and Type Ia Supernovae (SNe) and find that this can considerably enhance cosmological constraints in contrast to solely relying on GRBs. 
\vspace{2mm}
}

\maketitle
\flushbottom
\section{Introduction}
\hspace{0.45cm} Accelerated  expansion of the universe is  fairly  well established now \cite{1998AJ....116.1009R, 1999ApJ...517..565P, 2005ApJ...633..560E, 2004PhRvD..69j3501T}. The simplest model consistent with the observations is the spatially flat $\Lambda$CDM model, according to which the universe mostly consists of dark energy and dark matter, with baryonic matter being a small fraction. The  nature of the dark sector is still unknown and hence it becomes important to analyse the different models with combinations of observational datasets. For this, several groups have been trying to explore new probes in cosmology. GRBs are one such tool which has the  potential to explore the universe even at large redshift.\\

Gamma Ray Bursts (GRBs) are the most powerful explosions in the universe (with photon energies of the order of tens of keV to GeV) and are believed to form from the collapse of massive spinning stars \cite{2015PhR...561....1K}. On the basis of the burst duration, GRBs are classified in two categories: Short GRBs and Long GRBs. The short GRBs usually last for less than $2$ seconds while  long GRBs can last from $2$ seconds to several minutes  \cite{1993ApJ...413L.101K}. Though the exact mechanism behind these high energy explosions is not very clear, it is believed that the short and long GRBs form from the mergers of the binary neutron stars and core collapse of supernovae respectively \cite{2007PhR...442..166N, 2006ARA&A..44..507W}.\\ 
  
GRBs have been observed up to very high redshifts $( z\sim 9)$. Furthermore, the emission is  unaffected by the intervening dust. Due to these reasons, GRBs can be considered a good candidate to study the universe at high redshifts. Keeping this in mind, several energy-luminosity correlations for GRBs have been proposed in the literature.  The earliest correlation discovered by Amati et al., is known as the  ``Amati Relation" \cite{2002A&A...390...81A, 2006MNRAS.372..233A, 2008MNRAS.391..577A, 2009A&A...508..173A}. Other relations like the ``Ghirlanda relation", ``Yonetoku relation" and ``Liang-Zhang relation" were also proposed later that could make GRBs more suitable for standard candles  \cite{2004ApJ...616..331G, 2004ApJ...609..935Y, 2010A&A...511A..43G, 2005ApJ...633..611L,2022ApJ...924...97W}. 
Compared to these relations, the Amati relation has several advantages. First, it has a smaller scatter than the other relations, which means that it provides more accurate distance estimates. Second, it is easier to apply, as it only requires the measurement of two observables ($E_{\mathrm{iso}}$ and $E_\mathrm{p}$) that can be obtained from the prompt gamma-ray emission spectrum, rather than the detection of afterglow emission that is often challenging to observe. Finally, the Amati relation has been tested using a larger sample of GRBs than the other relations, making it a more reliable calibration tool.\\

The  Amati relation correlates the isotropic equivalent radiant energy $(E_{iso})$ with the spectral peak energy in the GRB rest frame $(E_p)$. Isotropic equivalent radiant energy is related to the luminosity distance, $(E_{iso}=4\pi d_L^2S(1+z)^{-1})$. Therefore, in order to calculate $E_{iso}$ from the observed GRB prompt fluence $(S)$, we need to first estimate the luminosity distance. However, due to the sparsity of  GRBs at low redshift, one has to assume a cosmological model. This means that the Amati relation is calibrated by assuming a cosmological model and a ``Circularity Problem" is  introduced if the same data is used to constrain the cosmological parameters. Apart from this, the Amati relation has also faced the extrinsic scatter problem. However, with better data, the scatter in the data has reduced and the relation has become more realistic.  \\
 
 To overcome the circularity issue in the Amati relation, various methods have been proposed in the literature. Some  authors have used the  GRB data with the $H(z)$ and BAO data to determine the cosmological and GRB correlation parameters simultaneously \cite{2019MNRAS.486L..46A, 2020MNRAS.499..391K}. In another  approach, several ancillary probes have been used to determine the luminosity distance in a model independent way. For example, authors have used Type Ia SNe data to calibrate the distance modulus of GRBs \cite{2022ApJ...931...50L, 2008ApJ...685..354L, 2008MNRAS.391L...1K, 2010PASJ...62.1495Y, 2022arXiv221102473N}. Observational Hubble data points have been used in literature to calibrate the Amati relation and approximate the cosmic evolution through a  Bezier parametric curve \cite{2023MNRAS.518.2247L, 2021MNRAS.501.3515M, 2019MNRAS.486L..46A, 2022arXiv220813700M,2021Galax...9...77L}. In a recent work, the angular diameter distances from galaxy clusters have been used  to calibrate the Amati relation at low redshifts  \cite{2022JCAP...10..069G}. \\ 

After the establishment of GRBs occurring at cosmological distances, various endeavors have been made to utilize GRB correlations to constrain cosmological parameters \cite{2005ApJ...633..611L,2010ApJ...714.1347S,2015GReGr..47..141L,2016MNRAS.455.2131L,2022MNRAS.514.1828D,2023MNRAS.518.2201D,2023arXiv230213887D}.
Such as an updated $E_{\mathrm{p}}-E_{\mathrm{iso}}$ correlation, using a cosmographic approach, has more recently been employed to put bound on cosmological parameters from a dataset of GRBs \cite{2017A&A...598A.112D}. Further, in their study, M. Demianski et al. (2019) employed the $E_{\mathrm{p}}-E_{\mathrm{iso}}$ correlation and incorporated potential changes in GRB observables to conclude that a scalar field with an exponential potential energy density describes a dynamical dark energy model that is consistent with calibrated GRB, Type Ia SNe, and H(z) data \cite{2019arXiv191108228D}.
In a recent study by N. Khadka et al (2021), they employed the Amati relation to obtain constraints on both the correlation and cosmological model parameters \cite{2021JCAP...09..042K}. 
S. Cao et al. (2022) utilized the A220 and A118 GRB samples \cite{2022MNRAS.512..439C,2022MNRAS.510.2928C}, along with the Dainotti-correlated GRB data sets compiled by J. P. Hu et al. (2021)  \cite{2021MNRAS.507..730H} and F. Y. Wang et al (2022) \cite{2022ApJ...924...97W}, to constrain cosmological model parameters by employing the Amati relation. On the other hand, people proposed improved versions of the Amati relation that incorporate a redshift-dependent term \cite{2022ApJ...931...50L,2022ApJ...935....7L}. To achieve this, they utilized a powerful statistical tool called copula. \\

In this context, there are two objectives of this work. In the \textbf{first} part, a constant Amati relation, with coefficients that do not depend on redshift, is calibrated without employing any cosmological model. For this, we use the Gaussian process to reconstruct the luminosity distance from the Hubble parameter measurements obtained from the differential ages of galaxies. Further, to check if the Amati relation parameters evolve with the GRBs' redshift, we divide the GRBs in five different redshift bins. For this, we use Old Astrophysical Objects (OAO) data and fit a third order polynomial with it. We then determine the luminosity distance using the derivative of the polynomial. It is important to note that in this way we study evolution of the Amati relation with redshift without adopting any background cosmological model.\\

After calibration of GRBs, in the \textbf{second} part, we use the constant Amati relation as well as the dynamical Amati relation with full GRBs data and determine the bounds on cosmological parameters of different models. In this work, we use three cosmological models: \textbf{i).} $\Lambda$CDM model, \textbf{ii).} $q(z)$ parametrizations, and \textbf{iii).} $z_t$CDM model. To study these models, we first adopt the  GRB data and then  perform a joint analysis with combined data of GRBs and Type Ia SNe.\\  


The paper is organized as follows. We describe the  data set and our  methodology in Section 2 and Section 3 respectively. Results are discussed in Section 4. Finally in Section 5, we discuss the final outcome of both parts along with the conclusions. 
\section{Observational Data sets}\label{dataset}

\subsection{Gamma Rays Bursts Data}
We use GRB data having $220$ data points (referred to as A220) in the redshift range, $0.0331 \leq z \leq 8.20$. This data is consolidated in literature (Tables 7 and 8 of ref. \cite{2021JCAP...09..042K}). In the data, corresponding to each sample source, the name of the GRB, its redshift, spectral peak energy in the rest frame ($E_p$), and measurement of the bolometric fluence ($S_{bolo}$) along with $1\sigma$ confidence level are mentioned. A220 is the union of two samples,  A118 ($0.3399 \leq z \leq 8.2$) and A102 ($0.0331 \leq z \leq 6.32$). The A118 data that has 118 long GRBs is further composed of two subsamples, i.e. 93 GRBs and 25 GRBs, collected from \cite{2016A&A...585A..68W} and \cite{2019ApJ...887...13F} respectively. A102 consist of 102 long GRBs taken from \cite{2016A&A...585A..68W, 2019MNRAS.486L..46A}.      

\subsection{Observational Hubble Data} 
We use the Hubble data to estimate the luminosity distance for GRBs calibration. In this work, we use H(z) data from the differential ages of passively evolving galaxies. 
For a  FLRW metric, $H(z)$ can be expressed as 

\begin{equation}\label{hz}
    H(z)=-\dfrac{1}{\left(1+z\right)}\dfrac{dz}{dt}
\end{equation}
 
\vspace{0.25cm}

It becomes essential to take proper care when using the differential age approach to estimate $H(z)$. With spectroscopy of extragalactic objects, redshift measurement is possible upto an accuracy of $\delta z/z <~ 10^{-3}$ but measurement of $dt$ is crucial. To estimate $dt$ various methods like full spectrum fitting, absorption feature analysis and calibration of specific spectroscopic features are used \cite{1994ApJS...95..107W,2011MNRAS.412.2183Tf, 2012JCAP...08..006M}. One of the important issues while using this approach is the degeneracy between the age-metallicity and age-star formation history. To overcome this, passively evolving red galaxies are used. It is assumed that the star formation in such galaxies has been quenched and hence their spectra are dominated by the older stellar population \cite{2003ApJ...593..622J}. The  measurement of differential age minimizes the systematics that could be there if we measure the absolute age. Measurement of $H(z)$ in this way does not rely on any cosmological model and is purely spectroscopic, making it a strong candidate to check the viability of, and assumptions behind, any cosmological model. Recently, Ma\~{g}an et. al (2018) compiled a dataset consisting of $31$ datapoints which were  measurements of $H(z)$ using differential ages of passively evolving galaxies \cite{2018MNRAS.476.1036M}. For our work, we have used 32 Hubble measurements that includes 31 from the Ma\~{g}an et. al (2018) compilation and one additional data point compiled by Borghi et al. (2022) at $z=0.75$ \cite{2022ApJ...928L...4B}. The redshift range of the data is $0.06\leq z\leq1.965$.

\subsection{Old Astrophysical Objects} 
\begin{figure}[h]
    \centering
    \includegraphics[width=0.6\textwidth]{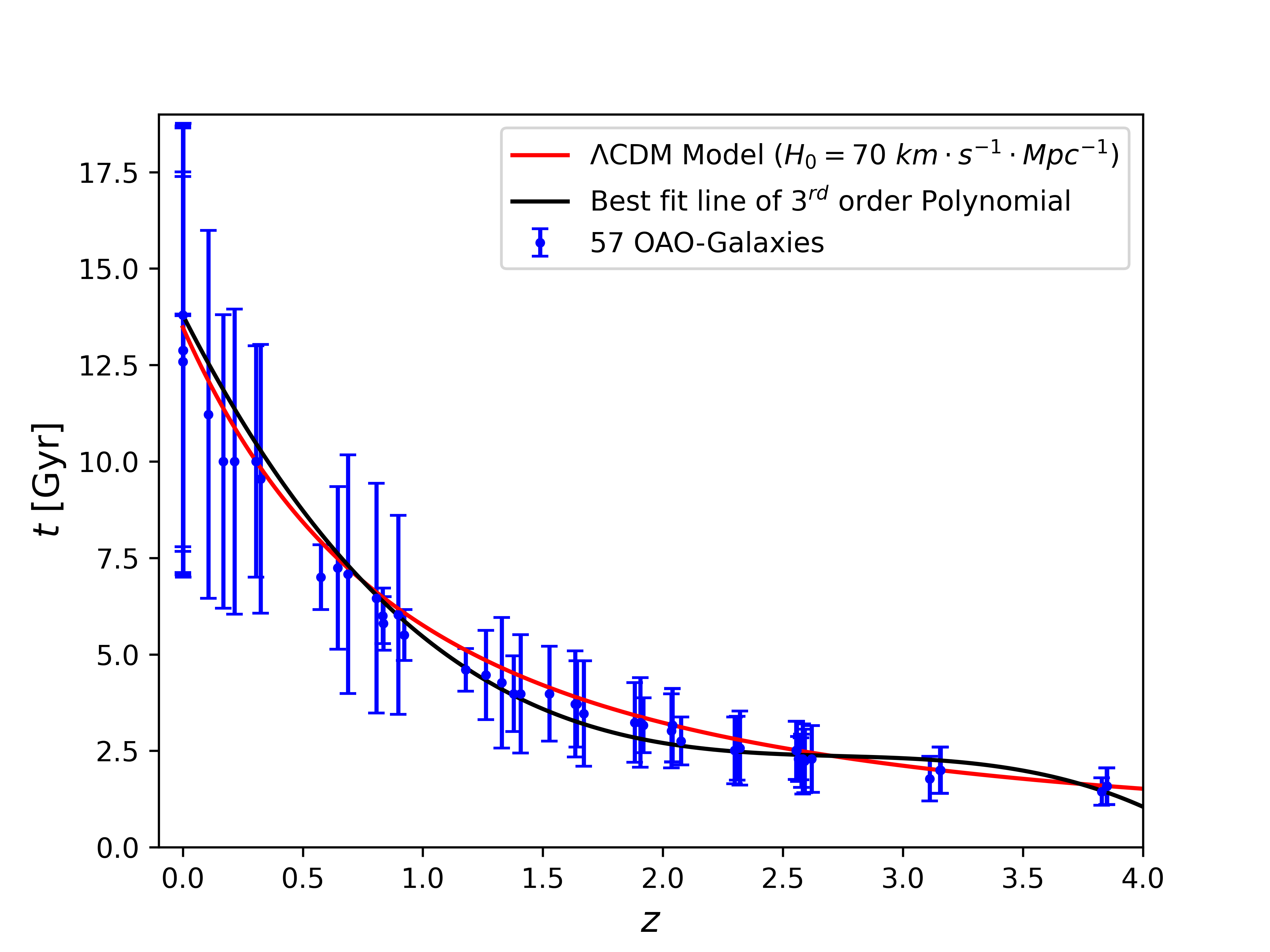}  
    \caption{Age-redshift plot for 57 datapoints of the  age of the galaxies from the Old Astrophysical Objects compilation in the redshift range, $0\leq z \leq 3.85$. The red and black lines represents $\Lambda$CDM model and the best fit for the third order polynomial respectively.}
    \label{t_vs_Z}
\end{figure}

Theoretically, the age of the universe at any redshift (z) is given as
$$t_u(z)=\int_z^\infty \frac{dz'}{(1+z')H(z')}$$
where $H(z)$ is the  Hubble parameter. The age of an old astronomical object $(t_i)$ at any redshift, $z_i$ can be given as the difference between the age of the universe when the object was born, i.e. $t_u(z_f)$ and the age of the universe at the redshift $z_i$, i.e. $t_u(z_i)$. 
We use 57 datapoints of the  age of the galaxies from the Old Astrophysical Objects compilation in the redshift range, $0\leq z \leq 3.85$ \cite{2022ApJ...928..165W}. Figure \ref{t_vs_Z} illustrates the age measurements of these 57 galaxies plotted with their respective redshifts. 

\subsection{Type Ia Supernovae Measurements} 
Type Ia supernovae are commonly regarded as standard candles. In the analysis of Type Ia supernovae, the distance modulus $(\mu_{\text{SNe}})$ is the standard observable quantity, which represents the difference between the apparent and absolute magnitude of the supernovae. For luminosity distance, we use the  Pantheon dataset of Type Ia supernovae, which includes 1048 data points within a redshift range of $0.01\leq z\leq2.26$ \cite{2018ApJ...859..101S}. The relationship that defines the observable measurement is expressed as

\begin{equation}
    \mu_{\mathrm{SNe}}(z)=m_{\mathrm{B}}(z)+\alpha \cdot X_{1}-\beta \cdot \mathcal{C}-M_{\mathrm{B}}.
\end{equation}
where $m_{\mathrm{B}}$ is the rest frame B-band peak magnitude, $M_B$ represents absolute B-band magnitude of a fiducial Type Ia SNe with $X_1 = 0$ and $C = 0$. Here $X_1$ and $C$ represent  the time stretch of light curve and supernova colour at maximum brightness respectively. In the Pantheon sample, the stretch-luminosity parameter $(\alpha)$ and the color-luminosity parameter $(\beta)$ have been calibrated to zero, resulting in only two remaining parameters. As a result, the distance modulus can be expressed as

\begin{equation}\label{mu_sn}
    \mu_{\mathrm{SNe}}(z)=m_{\mathrm{B}}(z)-M_{\mathrm{B}}.
\end{equation}

Furthermore, by obtaining the distance modulus, we can estimate both the luminosity distance $(d_L)$ and the uncertainty in the luminosity distance $(\sigma_{d_L})$ for each Type Ia supernova as 

\begin{equation} \label{eq_44}
    d_{\mathrm{L}}(z)=10^{\left(m_{\mathrm{B}}-M_{\mathrm{B}}-25\right) / 5}~~~(\mathrm{Mpc}), \quad \sigma_{d_{\mathrm{L}}}=\frac{\ln (10)}{5} d_{\mathrm{L}} \sigma_{m_{\mathrm{B}}}~~~(\mathrm{Mpc})
\end{equation}

As indicated by Eq. \eqref{eq_44}, determining the value of $M_B$ allows us to calculate the luminosity distance at a specific redshift. Recent research suggests that the luminosity (or absolute magnitude) of Type Ia supernovae does not evolve with redshift \cite{2019A&A...625A..15T,2022JCAP...01..053K}. So, it is generally accepted that the Type Ia supernovae sample is normally distributed with a mean absolute magnitude of $M_B = -19.22$ \cite{2020PhRvD.101j3517B}. However, in this work we take both $H_0$ and $M_B$ as free parameters since fixing one of them indirectly determines the other parameter.
 

\section{Methodology} \label{methodology}
\subsection{Gaussian Process} 
In this work, we need a model-independent estimate of the luminosity distance. For this, we first estimate the comoving distance from the Hubble parameter measurements in the redshift range $0<z<2$. We apply Gaussian Process (GP) to obtain a continuous smooth curve of $H(z)$ and after integrating the $H(z)$ values we get the corresponding comoving distances $(d_{C})$. GP is a well known hyper-parametric regression method which aims to reconstruct the shapes of physical functions from data without assuming a parametrized form of the function \cite{2006gpml.book.....R}. GP has been extensively used and because of its flexibility and simplicity, it is very useful for functional reconstructions. For example, from a set of measurements, we have $H(z)$ values and the uncertainty, i.e. $H(z_i)\pm \sigma_{H}$, where the value of $H(z_i)$ follows a Gaussian distribution at every point of $z_i$. Suppose, at an unknown point $z^\prime$, we want to estimate the value of the function. For this, we have a covariance or kernel function $k(z,z^\prime)$ which indicates that the value of the function at $z$ is not independent of its value at $z^\prime$ but instead the values are correlated by the kernel function.\\ 

Gaussian Process is a non-parametric technique because it only depends on the choice of the covariance function and not on model parameters or any assumed functional form. The covariance function often depends on the separation $|z-z^\prime|$ between the points. In this analysis, we consider the Squared Exponential or Gaussian kernel function \cite{2012JCAP...08..006M,2021EPJC...81..892O} since this function has the characteristic of being infinitely differentiable, which is important for reconstructing a derivative of a function. The Squared Exponential kernel function is 
 
\begin{equation}
    k\left(z, z^{\prime}\right)=\sigma_f^2 \exp \left(-\frac{\left(z-z^{\prime}\right)^2}{2 \ell^2}\right)
\end{equation}

where $\sigma_f$ and $\ell$ are the GP hyperparameters which basically regulate the correlation-strength of the function value and the length scale of the correlation in $z$ respectively. Using the observed data, one can estimate the value of $\sigma_f$ and $\ell$ parameters by minimizing a log marginal likelihood function. For maximization, we use flat priors for the $\sigma_f$ and $\ell$ parameters of the kernel function.\\

Once we get the reconstructed $H(z)$ in the required redshift range,  $0<z<2$, we use the Simpson $\frac{3}{8}$ method for numerical integration of $1/H(z)$ to obtain continuous values of $d_C$ in the same redshift range as  

\begin{equation}\label{gp1_1}
   {d_C}(z)=\displaystyle\int_0^z\dfrac{\Tilde{c} dz^\prime}{H(z^\prime)}  
\end{equation}
\hspace{0.35cm} where $\Tilde{c}$ is speed of light.\\

Finally, to obtain luminosity distance we use $d_L=(1+z)d_C$. The corresponding uncertainties are obtained by propagating the error obtained in $d_C$ using Gaussian Process as shown in Figure \ref{figure1}. The  $d_L$ obtained by this method is then used with $S_{bolo}$ given in the GRB data to calculate $E_{iso}$. It is important to note that only $118$ data points lying in the redshift range $0< z<2$ of A220 data are considered for this purpose as the Hubble data used here also lie in this redshift range. 
\\

\subsection{Parameter Estimation}
In this section, we basically calibrate the GRB Amati relation via Gaussian process. Using this calibration, we derive distance moduli of GRBs at high redshifts and hence put constraints on different cosmological model parameters. Therefore, first we calibrate the Amati relation (the $E_{\text{p}}-E_{\text{iso}}$ correlation) of GRBs in a cosmology-independent way.

\subsubsection{GRBs calibration with the Amati relation}
For this analysis, we use a linear regression relation using the logarithms of $E_{\text{iso}}$ and $E_{\text{P}}$. This relation is generally referred to as the Amati relation which is basically a correlation between isotropic equivalent energy ($E_{\text{iso}}$) and spectrum peak energy in the comoving frame ($E_{\text{P}}$). The Amati relation can be parametrized as 

\begin{equation}\label{GRB_1}
    \log\left[\dfrac{E_{\text P}}{1~\text{keV}}\right]=m\log\left[\dfrac{E_{\text{iso}}}{1~\text{erg}}\right]+c
\end{equation}

where, 
\begin{equation}\label{GRB_2}
    E_{\text{iso}}=\dfrac{4\pi d_L^2S_{\text{bolo}}}{\left(1+z\right)}
\end{equation} \\

Here $S_{\text{bolo}}$ is the bolometric fluence and $d_L$ is luminosity distance which we estimate from the comoving distance as $d_L=d_{C}(1+z)$. The factor $(1+z)$ accounts for the cosmological time dilation effect. The peak energy in the comoving frame, $E_P$, is related to the observed peak energy $E_{P}^\text{obs}$ by


\begin{equation}\label{GRB_3}
    E_\text{P}=E_{\text{P}}^{\text{obs}}\left(1+z\right)
\end{equation}

Defining,

\begin{equation}\label{GRB_4}
    y\equiv\log\left[\dfrac{E_{\text P}}{1~\text{keV}}\right],~~~~~~~x\equiv\log\left[\dfrac{E_{\text{iso}}}{1~\text{erg}}\right]
\end{equation}

we can rewrite Eq. (\ref{GRB_1}) as

\begin{equation}\label{GRB_5}
    y=mx+c
\end{equation}

The associated uncertainties with $y$ and $x$ are given as

\begin{equation}\label{GRB_6}
    \sigma_y=\dfrac{1}{\ln(10)}\left(\dfrac{\sigma_{_{E_{\text P}}}}{E_{\text P}}\right),~~~~~~~\sigma_x=\dfrac{1}{\ln(10)}\left(\dfrac{\sigma_{_{E_{\text iso}}}}{E_{\text iso}}\right)
\end{equation}

In the Amati relation mentioned above, we have two parameters, namely the slope $(m)$ and the  intercept $(c)$. These parameters can be estimated by directly fitting  Eq. \eqref{GRB_5} with the observed GRB data. In this analysis, these parameters are determined by maximizing the likelihood ($\mathcal{L}$) defined as

\begin{equation}\label{GRB_7}
    -2 \ln \mathcal{L}=\sum_i \ln 2 \pi \sigma_i^2+\sum_i \frac{\left[y_i-\left(m x_i+c\right)\right]^2}{\sigma_i^2}
\end{equation}

where $\sigma_i^2={\sigma_{y_i}^2+m^2 \sigma_{x_i}^2}+\sigma_s^2$ and $\sigma_s$ denotes the intrinsic scatter which specifies the tightness of the Amati relation.

\subsubsection{Cosmological Parameters}
After obtaining the values of the Amati parameters from the low-redshift GRBs, we extrapolate this relation to obtain $E_{\text{iso}}$ at high redshifts. Therefore, using Eq. \eqref{GRB_2} we estimate the luminosity distance or distance modulus as

\begin{equation}
    \mu_{\text{GRB}}=\dfrac{5}{2}\left[\log\left(\dfrac{E_{\text{iso}}}{1~\text{erg}}\right)-\log\left(\dfrac{4\pi}{\left(1+z\right)}\right)-\log\left(\dfrac{S_{\text{bolo}}}{1~\text{erg/cm}^2}\right)\right]-97.45
\end{equation}
 
And uncertainty in $\mu_{\text{GRB}}$ is

\begin{equation}
    \sigma_{\mu_{\text{GRB}}}^2=\dfrac{25}{4}\left[\left( \sigma_{\log \left(\frac{E_{\text {iso }}}{1 \text { erg }}\right)}\right)^2+\left(\frac{1}{\ln 10} \frac{\sigma_{S_{\text {bolo }}}}{S_{\text {bolo }}}\right)^2\right]
\end{equation}
where $\sigma_{\log \left(\frac{E_{\text {iso }}}{1 \text { erg }}\right)}$ can be calculated using the error propagation in Eq. \ref{GRB_1}.
\vspace{3mm}\\
Based on the distance moduli of the observed data points, we study the kinematics of our universe. The advantage of using this approach lies in the fact that it allows one to assess the direct evidence of the late-time acceleration in the universe without relying on its matter-energy content \cite{turner2002type}. We consider three different cosmological models, i.e. $\Lambda$CDM model, $q(z)$ parametrizations and $z_t$CDM model. These models depend on a combination of different cosmological parameters. In this analysis, the parameters are collectively denoted by the symbol $\mathbb{P}$. 
\begin{itemize}
    \item \textbf{$\Lambda$CDM Model:} 
    In this model, the universe is spatially flat and  the matter density (baryonic and dark matter), as determined by the Planck result, contributes approximately $31.47 \pm 0.74$\% to the total energy density of the universe. The rest of the contribution comes from dark energy \cite{2020A&A...641A...6P}. This model is in concordance with most of the observations and hence it is widely accepted. In a flat universe, the Hubble parameter for the  $\Lambda$CDM model is given as
    $$H(z,\mathbb{P})=H_0\sqrt{\Omega_{m0}(1+z)^3+(1-\Omega_{m0})}$$
    
    \item \textbf{$q(z)$ Parametrizations:} In this analysis, we consider three different parametrizations of $q(z)$, one with one parameter and two with two parameters. Corresponding to each parametrization, one can determine expression for Hubble parameter using the following equation $$H(z,\mathbb{P})=H_0\exp{\left[\displaystyle\int_0^z\dfrac{1+q(x)}{1+x}dx\right]}$$.\\ 
    The three parametrizations are :
    \begin{itemize}
        \item \textbf{P1:} $q(z)=q_0$; This parametrization is motivated by the fact that a value less than $0$ is indicative of recent acceleration without reference to any particular dark energy model \cite{2002ApJ...569...18T}.
        
        \item \textbf{P2:} $q(z)=q_0+q_1z$; This continuous and smooth linear parametrization gives the value of deceleration parameter at $z=0$ while at higher redshifts it diverges \cite{Riess_2004}.
        
        \item \textbf{P3:} $q(z)=q_0+q_1\dfrac{z}{1+z}$; This parametrization is better than the  P2  parametrization as it behaves well at high redshifts while the linear approach diverges in  the distant past \cite{2008MPLA...23.1939X}.
        
    \end{itemize}
    
    \item \textbf{$z_t$CDM Model:} If we use the condition ${\Ddot{a}}=0$ in the $\Lambda$CDM model, we obtain an expression for the transition redshift, $$z_t=\left(\frac{2 \Omega_{\Lambda 0}}{\Omega_{m0}} \right)^{1/3}-1 $$ Substituting for $z_t$ in the flat $\Lambda$CDM model gives us \cite{2022arXiv221204751D, Velasquez-Toribio:2020had}. 
    
    $$H(z,\mathbb{P})=H_0\sqrt{\dfrac{(1+z)^3}{0.5(1+z_t)^3+1}+\dfrac{(1+z_t)^3}{(1+z_t)^3+2}}$$
\end{itemize}
We have assumed a spatially flat cosmology as the background for all the models mentioned above. Considering these Hubble functions,  the theoretical value of the luminosity distance can be obtained from

\begin{equation}\label{eq_dl_cosmo}
    d_L(z,\mathbb{P})=\Tilde{c}(1+z)\displaystyle\int_0^z \dfrac{1}{H(z^\prime;\mathbb{P})}dz^\prime
\end{equation}
where 
$\Tilde{c}$ is the velocity of light. Using Eq. \eqref{eq_dl_cosmo},  we finally  estimate the theoretical distance modulus as 
\begin{equation}\label{mu_th}
    \mu_{th}(z,\mathbb{P})=5\log\left(\dfrac{d_L(z;\mathbb{P})}{1~\text{Mpc}}\right)+25
\end{equation} 

By utilizing GRB data at $0.03<z<8.2$, we fit the cosmological parameters  using the $\chi^2$ minimization technique as

\begin{equation}\label{chi_grb}
    \chi_{\mathrm{GRB}}^2=\sum_{i=1}^{N_{\mathrm{G}}}\left[\frac{\mu_{\mathrm{GRB}}\left(z_i\right)-\mu_{\mathrm{th}}\left(z_i ; \mathbb{P}\right)}{\sigma_{\mu_{\mathrm{GRB}}}(z_i)}\right]^2
\end{equation}
Here $N_\mathrm{G}=220$ denotes the total number of GRB datapoints. 

\vspace{3mm}
In addition to incorporating the GRBs, we also include the Pantheon database of Type Ia SNe. This data  is useful for obtaining bounds on cosmological parameters and distinguishing between models of dark energy. Using Eqs. (\ref{mu_sn}, \ref{mu_th}) the $\chi^2$ for this database is defined as

\begin{equation}\label{chi_sne}
    \chi_{\mathrm{SNe}}^2=\sum_{i=1}^{N_\mathrm{S}}\left[\frac{\mu_{\mathrm{SNe}}\left(z_i\right)-\mu_{\mathrm{th}}\left(z_i ; \mathbb{P}\right)}{\sigma_{\mu_{\mathrm{SNe}}}(z_i)}\right]^2
\end{equation}
where, $N_\mathrm{S}$ denotes the total number of Type Ia SNe.\\

Thus, the total $\chi^2_\mathrm{T}$ of GRBs and Type Ia SNe is

\begin{equation}
    \chi_{\mathrm{T}}^2=\chi_{\mathrm{GRB}}^2+\chi_{\mathrm{SNe}}^2
\end{equation}

\vspace{7mm}
In this analysis, we employ the Markov Chain Monte Carlo (MCMC) analysis using \textbf{ {\texttt{emcee}}}, a Python package introduced by Foreman-Mackey and colleagues in 2013 \cite{emcee}. After executing the MCMC method to obtain the best fit for all parameters, we determine the confidence levels and their corresponding $1\sigma, ~2\sigma,$ and $3\sigma$ uncertainties using the \textbf{ {\texttt{corner}}} Python package \cite{corner}. For this work, we adopt a wide, uniform prior for all parameters as shown in Table \ref{table_prior}.

\begin{table}
\begin{center}
\renewcommand{\arraystretch}{1.5}
    \begin{tabular}{ |c|c| } 
\hline
Parameters & Prior Range   \\
\hline
\hline
 m & $\mathcal{U}[-1,1]$   \\ 
\hline
c& $\mathcal{U}[-30,10]$\\ 
\hline
$\sigma_s$& $\mathcal{U}[0,2]$ \\ 
\hline
$H_0$ & $\mathcal{U}[40, 100]$ \\
\hline
$\Omega_{m0}$ & $\mathcal{U}[0,1]$ \\
\hline
$q_0$ & $\mathcal{U}[-2,2]$ \\
\hline
$q_1$ & $\mathcal{U}[-4,4]$ \\
\hline
$z_t$ & $\mathcal{U}[0,2]$ \\
\hline
$M_B$ & $\mathcal{U}[-24,-10]$\\
\hline
\end{tabular}
\caption{The prior range of Amati correlation parameters as well as cosmological model parameters sampled in MCMCs. The notation $\mathcal{U}[a, b]$ refers to a uniform distribution  over the interval $[a, b]$. }
\label{table_prior} 
\end{center}
\end{table}
%
\section{Results}\label{results}
In this work, our aim is to calibrate the Amati relation for Gamma Rays Bursts in a model independent way. This paper is divided into two parts. In the first part of paper, we calibrate the $E_\mathrm{p}-E_\mathrm{iso}$ correlation of gamma-ray bursts (GRBs), known as the Amati relation, that is not based on any specific cosmological model.
In the second part, our focus shifts to investigate the cosmological models using the observed GRBs data points as well as supernovae data.

\subsection{Calibration of Gamma Rays Bursts}
Our aim is to get bounds on the Amati relation parameters ($m$ and $c$) and the intrinsic scatter ($\sigma_s$) by maximising the likelihood. For this we require $E_{\text{P}}$ and $E_{\text{iso}}$, where $E_{\text{P}}$ can be obtained directly from the GRB data. But to estimate $E_{\text{iso}}$, we need to determine luminosity distance $(d_L)$ corresponding to each redshift of GRBs.
 
\subsubsection{Constant Amati Relation Calibration}
\begin{figure}[H]  
    \centering
    \includegraphics[width=0.9\textwidth]{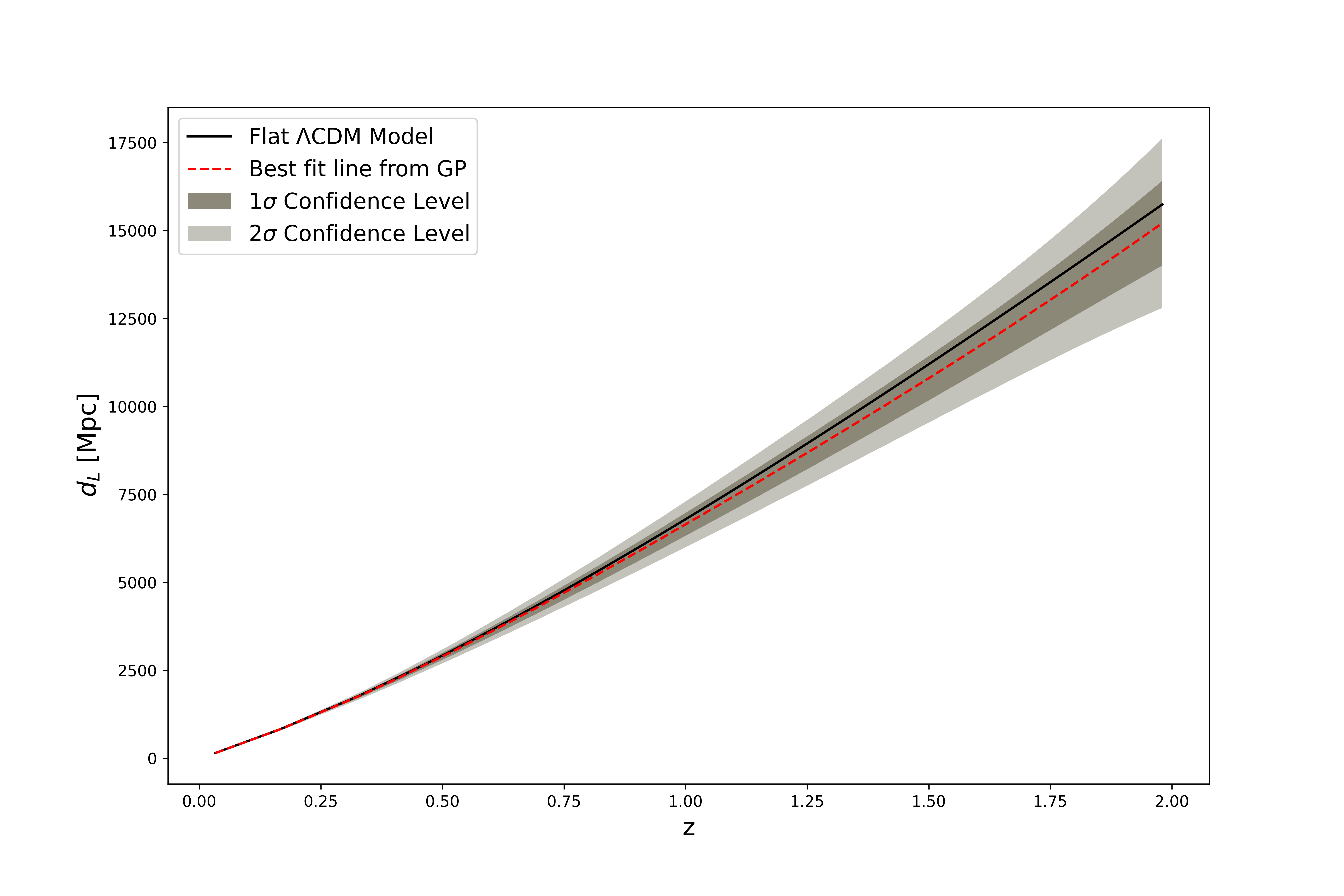}  
    \caption{This plot represents the reconstructed values of luminosity distance, $d_L$ in the redshift range $0 < z < 2$.
    }
     \label{figure1}
\end{figure}

We used a non-parametric method (Gaussian Process) to reconstruct luminosity distance from the observational Hubble data. The reconstructed $d_L$ vs $\textit{z}$ curve is shown in Figure \ref{figure1}. The red dashed curve represents the reconstructed $d_L$ line while the dark and light grey colors are $1\sigma$ and $2\sigma$ confidence regions respectively. The black line shows variation of the luminosity distance with redshift for a flat $\Lambda$CDM model with $\Omega_{m0}=0.3$.

\begin{figure}[ht]  
    \centering
     \includegraphics[width=90mm]{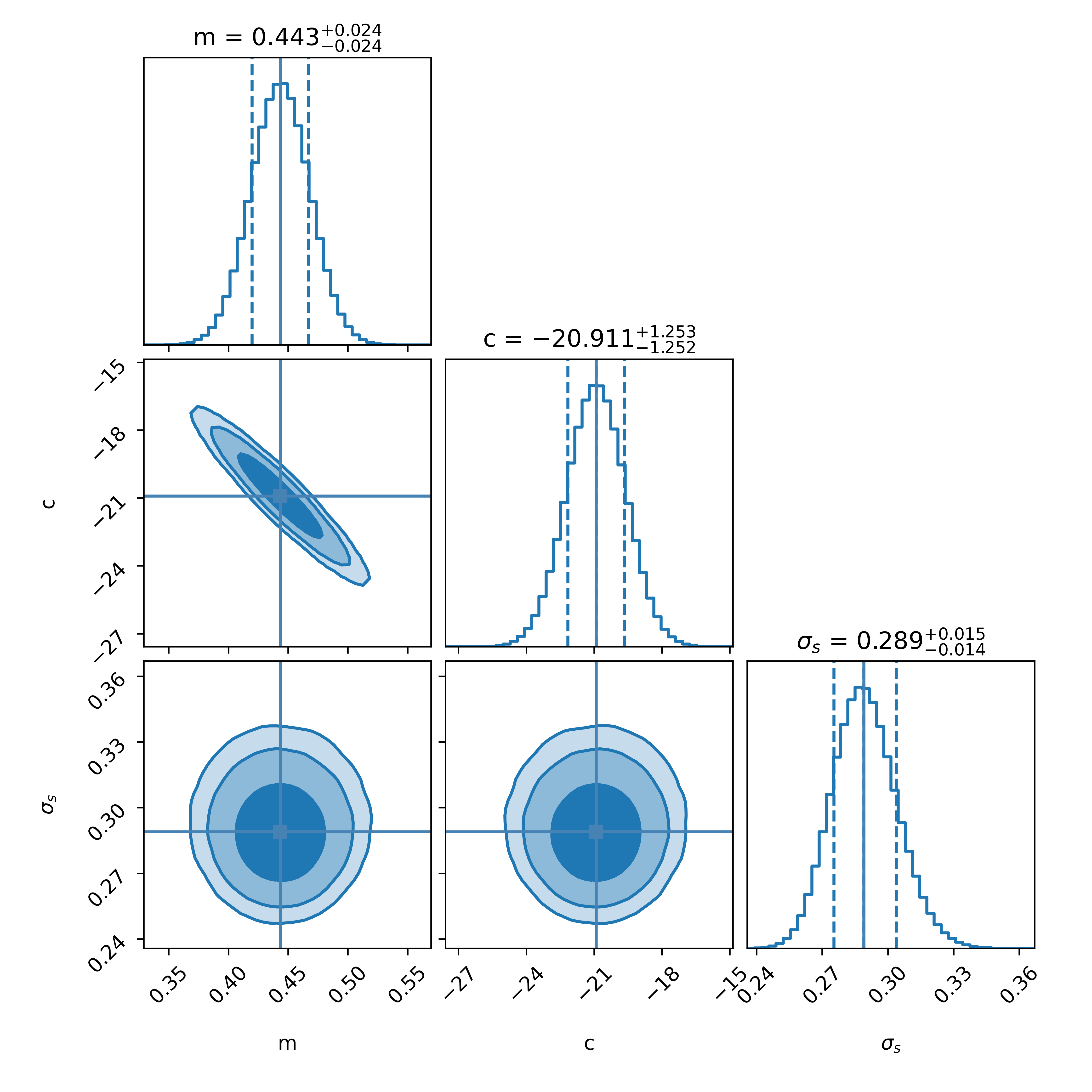}   
     \caption{ 68\%, 95\% and 99\% contours of the Amati relation parameter, i.e. m, c and $\sigma_s$ for the subset of A220 GRB data upto $z<2$.}
     \label{figure2}
\end{figure}

\begin{table}[H]
\begin{center}
\renewcommand{\arraystretch}{1.5}
    \begin{tabular}{ |c|c|c|c| } 
\hline
Parameters & Galaxy cluster \cite{2022JCAP...10..069G}  &  $H(z)$, our results  \\
\hline
 m & $0.44^{+0.07}_{-0.09}$   &$0.443^{+0.024}_{-0.024}$  \\ 
\hline
c& $-20.22^{+4.52}_{-3.76}$   &  $-20.911^{+1.253}_{-1.252}$\\ 
\hline
$\sigma_s$& $0.45^{+0.091}_{-0.066}$   & $0.289^{+0.015}_{-0.014}$  \\ 
\hline
\end{tabular}
\caption{The best fit values of $m$, $c$ and $\sigma_s$ with 68\% confidence level obtained using GRB dataset.}
\label{table1}
\end{center}
\end{table}

Based on this analysis, the error bars on Amati relation parameter reduces as compared to the results obtained by using the galaxy cluster data \cite{2022JCAP...10..069G}. This suggests that the analysis used here is more accurate and precise in constraining the desired parameters (see Table \ref{table1} and Figure \ref{figure2}). \\

Note that there have been different forms of the Amati relation mentioned in the literature \cite{2020MNRAS.499..391K,2022arXiv221102473N,2016MNRAS.455.2131L}. To test the validity of these alternative forms, we have performed an analysis using their Amati relation and we find that the our results are consistent with their published results.


\subsubsection{Dynamical Amati Relation Calibration}

Further, in order to examine whether the Amati relation changes over time or redshift, we use technique carried by \cite{2007MNRAS.379L..55L} and divide the set of GRBs till redshift 3.85 into five groups based on their redshifts from low to high. Each bin has 40 GRBs as mentioned below.

\begin{itemize}
    \item \textbf{B1:} $0.0331\leq z\leq 0.8090$ $\Rightarrow$ ${<}\textit{z}{>}=0.569$
    \item \textbf{B2:} $0.8090<z\leq1.4050$ $\Rightarrow$ ${<}\textit{z}{>}=1.109$
    \item \textbf{B3:} $1.4050<z\leq2.0100$ $\Rightarrow$ ${<}\textit{z}{>}=1.6653$
    \item \textbf{B4:} $2.0100<z\leq2.5910$ $\Rightarrow$ ${<}\textit{z}{>}=2.265$
    \item \textbf{B5:} $2.5910<z\leq3.8000$ $\Rightarrow$ ${<}\textit{z}{>}=3.122$
\end{itemize}

We then apply Eq. \eqref{GRB_1} 
to each bin to find the best fit value of Amati relation parameters. It is important to note that in order to use Eq. \eqref{GRB_1}, we need the luminosity distance. We determine $d_L$ in a model independent way by using the Old Astrophysical Objects data. We parameterize the age of old objects as a third order polynomial and fit it with the observed data.

\begin{equation}\label{oao}
   t(z)=A+Bz+Cz^2+Dz^3
\end{equation}

The obtained best fit values of A, B, C and D are $13.798^{+0.015}_{-0.015}$, $-12.203^{+0.391}_{-0.391}$, $4.402^{+0.282}_{-0.282}$ and $D=-0.537^{+0.048}_{-0.048}$ respectively. Then we write comoving distance in terms of these constants (see Eq. \eqref{dcom}) and calculate $d_L$ from $d_{co}$.

\begin{equation}\label{derioao}
    H(z)=-\frac{1}{(1+z)(B+2Cz+3Dz^2)}
\end{equation}
Hence, the comoving distance is;
\begin{equation}\label{dcom}
   d_{co}=-\Tilde{c}\left[B\left(z+\frac{z^2}{2}\right)+C\left(z^2+\frac{2z^3}{3}\right)+D\left(z^3+\frac{3z^4}{4}\right)\right]
\end{equation}

The best fit values of the Amati relation parameters for each bin are shown in Table \ref{tab_a_b_binned}.
 \begin{table}[H]
        \begin{center}
        \renewcommand{\arraystretch}{1.5}
            \begin{tabular}{ |c|c|c|c| } 
                \hline
                {Bin} & $m$ & $c$ & $\sigma_s$ \\
                \hline
                \textbf{B1} & $0.376^{+0.043}_{-0.042}$  & $-17.467^{+2.222}_{-2.229}$ & $0.292^{+0.026}_{-0.023}$ \\
                \hline
                \textbf{B2} & $0.473^{+0.046}_{-0.046}$ & $-22.457^{+2.439}_{-2.425}$ & $0.299^{+0.027}_{-0.024}$  \\
                 \hline
                \textbf{B3} & $0.390^{+0.039}_{-0.039}$  & $-18.101^{+2.064}_{-2.072}$  & $0.235^{+0.024}_{-0.021}$  \\
                 \hline
                \textbf{B4} & $0.375^{+0.057}_{-0.058}$  & $-17.215^{+3.074}_{-3.064}$ & $0.293^{+0.029}_{-0.026}$ \\
                 \hline
                \textbf{B5} &$0.500^{+0.052}_{-0.052}$ & $-23.983^{+2.774}_{-2.776}$ & $0.137^{+0.044}_{-0.052}$\\
                 \hline 
            \end{tabular}
        \caption{The best fit values of $m$, $c$ and $\sigma_s$ with 68\% confidence.}
        \label{tab_a_b_binned}
        \end{center}
    \end{table}


Now, using the best fit values of $m$, we do a linear fit to $m$ versus ${<}\textit{z}{>}$ and find the equation as

\begin{equation}
    m=0.368(\pm0.029)+0.031(\pm0.015)z
\end{equation}

Similarly, we find an equation for a linear fit to $c$ versus ${<}\textit{z}{>}$ as 

\begin{equation}
    c=-17.327(\pm1.620)-1.439(\pm0.902)z
\end{equation}

\begin{figure}[H]
\centering
\begin{minipage}{.475\textwidth}
  \centering
  \includegraphics[width=1.2\linewidth]{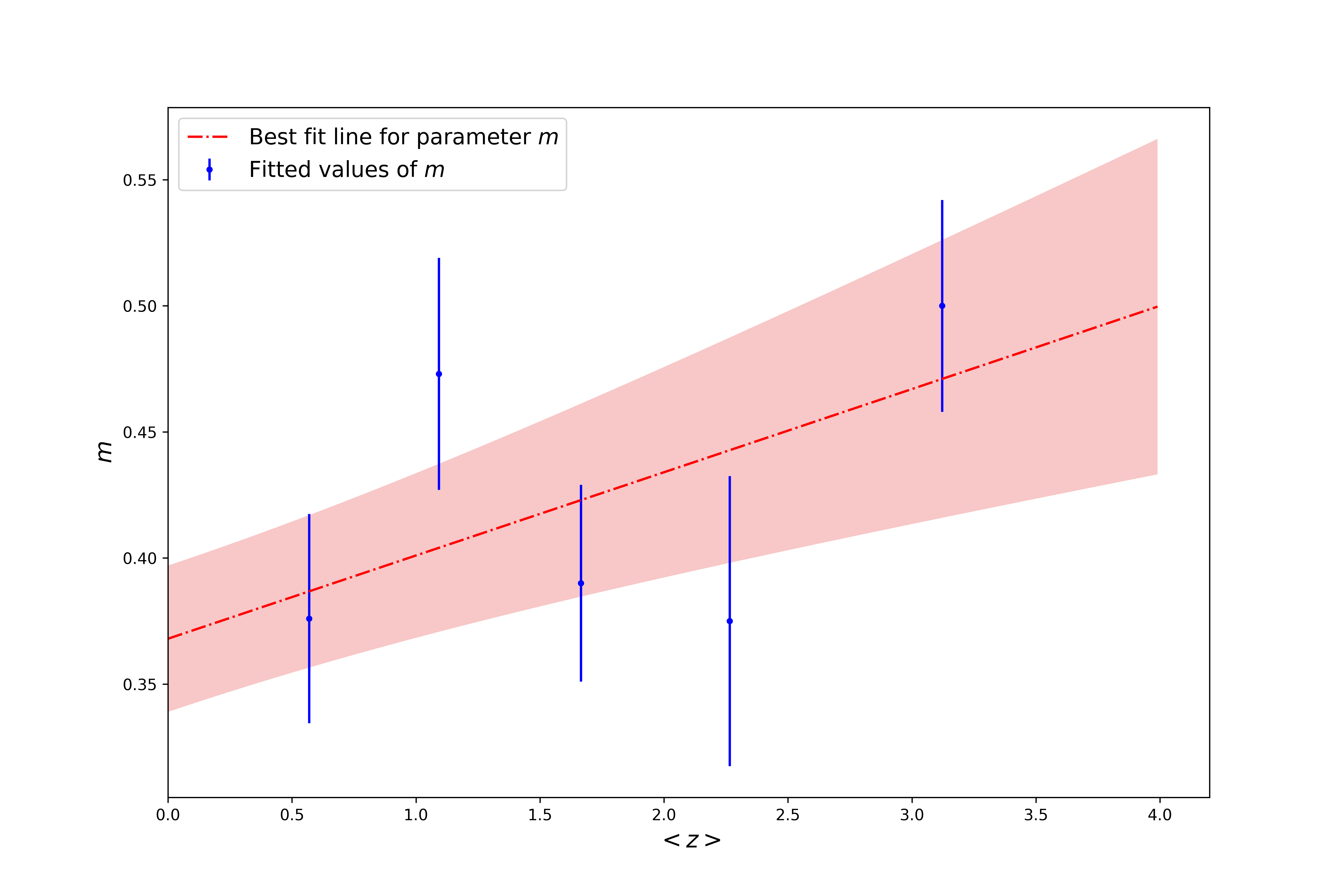}
\end{minipage}%
\hfill
\begin{minipage}{.475\textwidth} 
  \centering
  \includegraphics[width=1.2\linewidth]{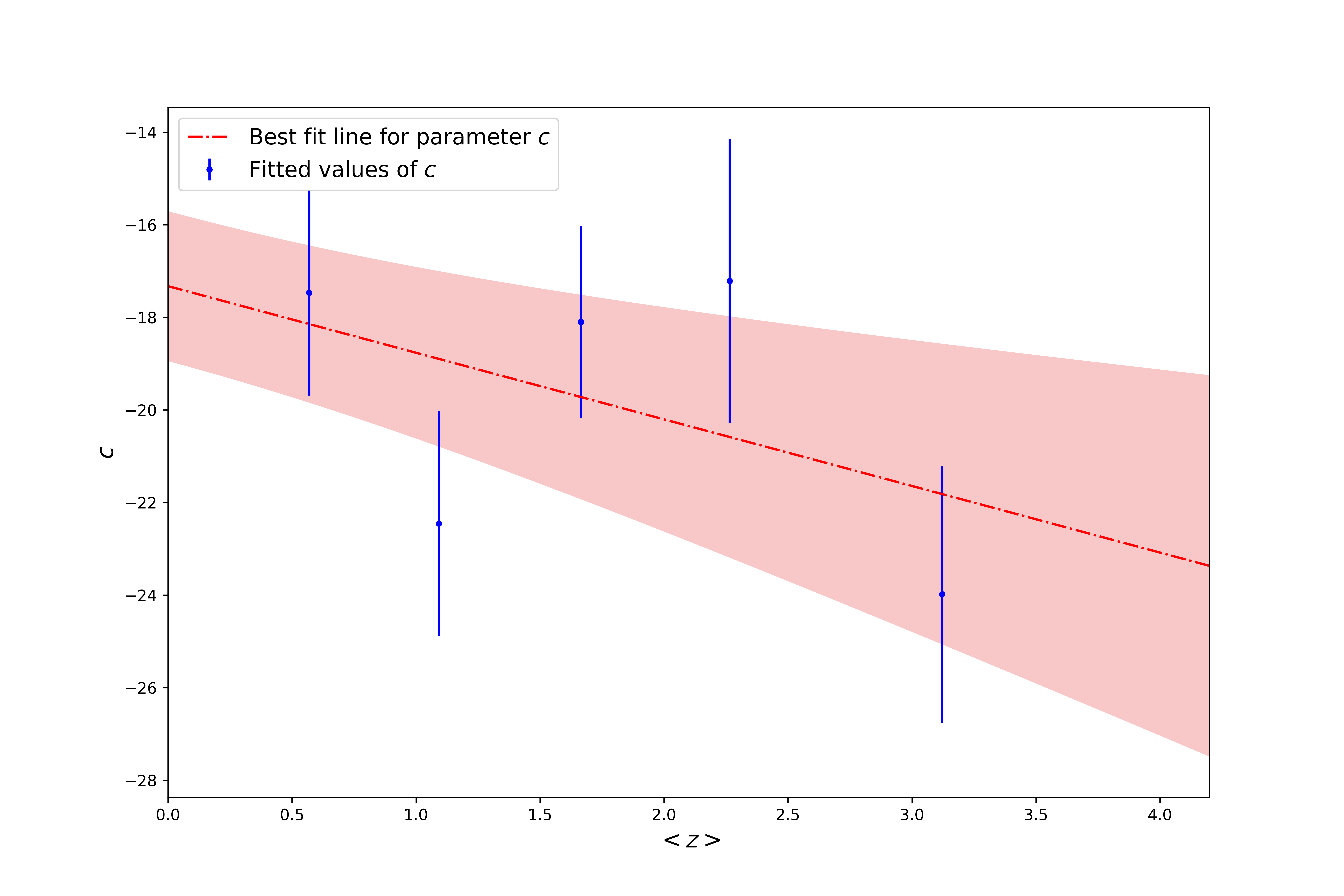}
\end{minipage}
\caption{\textbf{Left panel:} The variation of $m$ with mean redshift, \textbf{Right panel:} The variation of $c$ with mean redshift. The dashed-dotted red line represents the best fit line and 68\% confidence level is shown in pink color band.}
    \label{fig_m_c_best_fit}
\end{figure}

Variation of Amati relation parameters $(m~\&~c)$ with mean redshift is shown in Figure \ref{fig_m_c_best_fit}. 

From Figure \ref{fig_m_c_best_fit}, it seems that $m$ and $c$ do evolve with redshift.

\subsection{Constraints on Cosmological Parameters}

\subsubsection{Fit with Gamma Rays Bursts Dataset}

We study the cosmological models both with constant Amati relation parameters $(m,~c)$ and dynamical Amati relation parameters $(m(z),~c(z))$.

\subsubsection*{1. \underline{$\mathbf{\Lambda}$CDM Model:}}
The best fit values of $H_0$ and $\Omega_{m0}$ obtained for $\Lambda$CDM model are given in Table \ref{tab_csmly_lcdm_grbs} .

\begin{table}[H]
\begin{center}
\renewcommand{\arraystretch}{1.5}
    \begin{tabular}{ |c|c|c|c| } 
\hline
Parameters & Constant $m~\&~c$  & Dynamical $m~\&~c$ & Figure  \\
\hline
 $H_{0}$ & $67.188^{+16.980}_{-11.809}$   & $65.449^{+17.504}_{-11.003}$  &\multirow{2}{*}{Figure \ref{csmlgy_lcdm_grbs}}  \\ 
 \cline{1-3}
 $\Omega_{m0}$ & $0.314^{+0.357}_{-0.213}$   &$0.411^{+0.359}_{-0.273}$ &  \\ 
\hline
\end{tabular}
\caption{The best fit values of $H_{0}$ and $\Omega_{m0}$ with 68\% confidence level obtained for $\Lambda$CDM model using GRB dataset.}
\label{tab_csmly_lcdm_grbs}
\end{center}
\end{table}

\begin{figure}[H]
\centering
\begin{minipage}{.475\textwidth}
  \centering
  \includegraphics[width=1.1\linewidth]{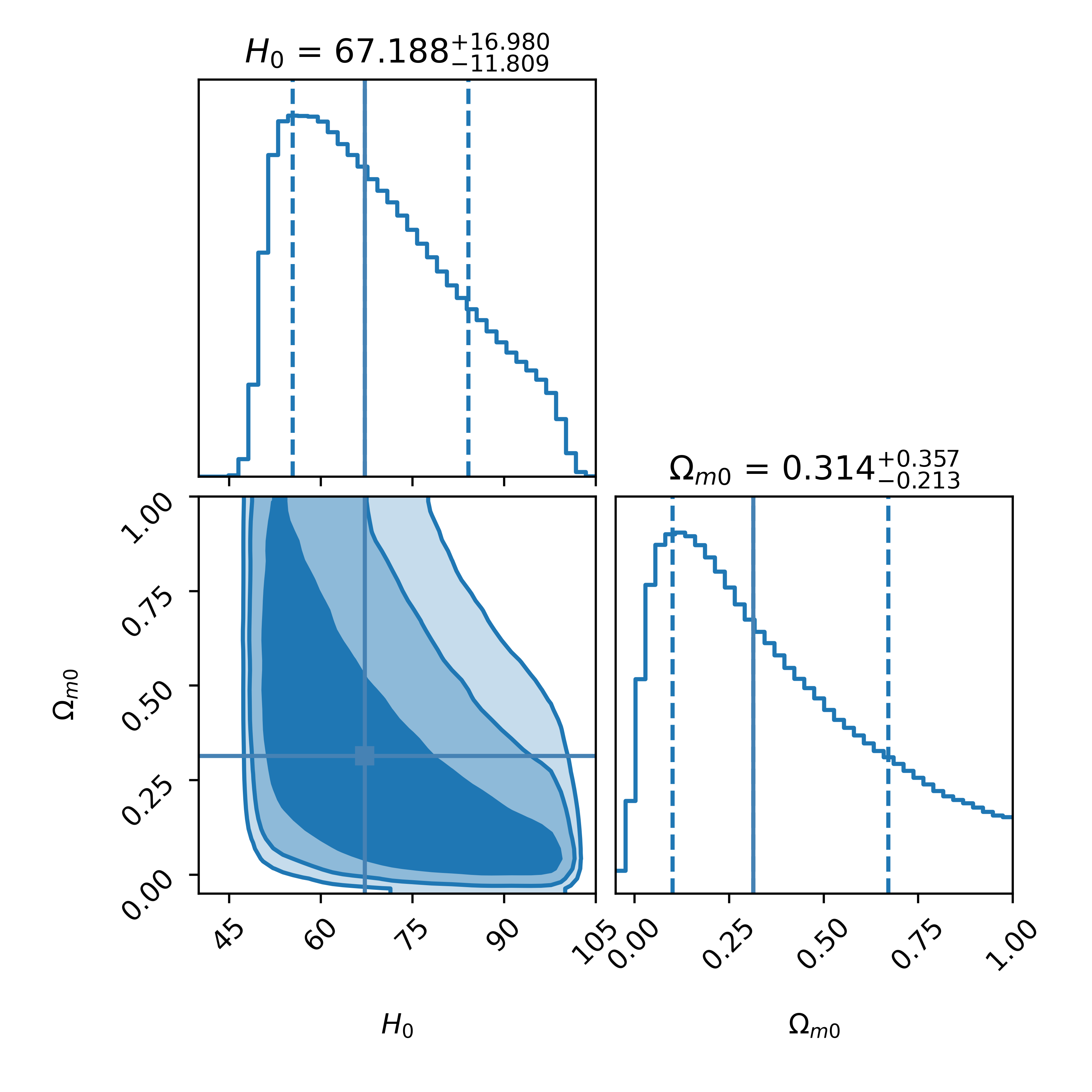}
\end{minipage}%
\hfill
\begin{minipage}{.475\textwidth}
  \centering
  \includegraphics[width=1.2\linewidth]{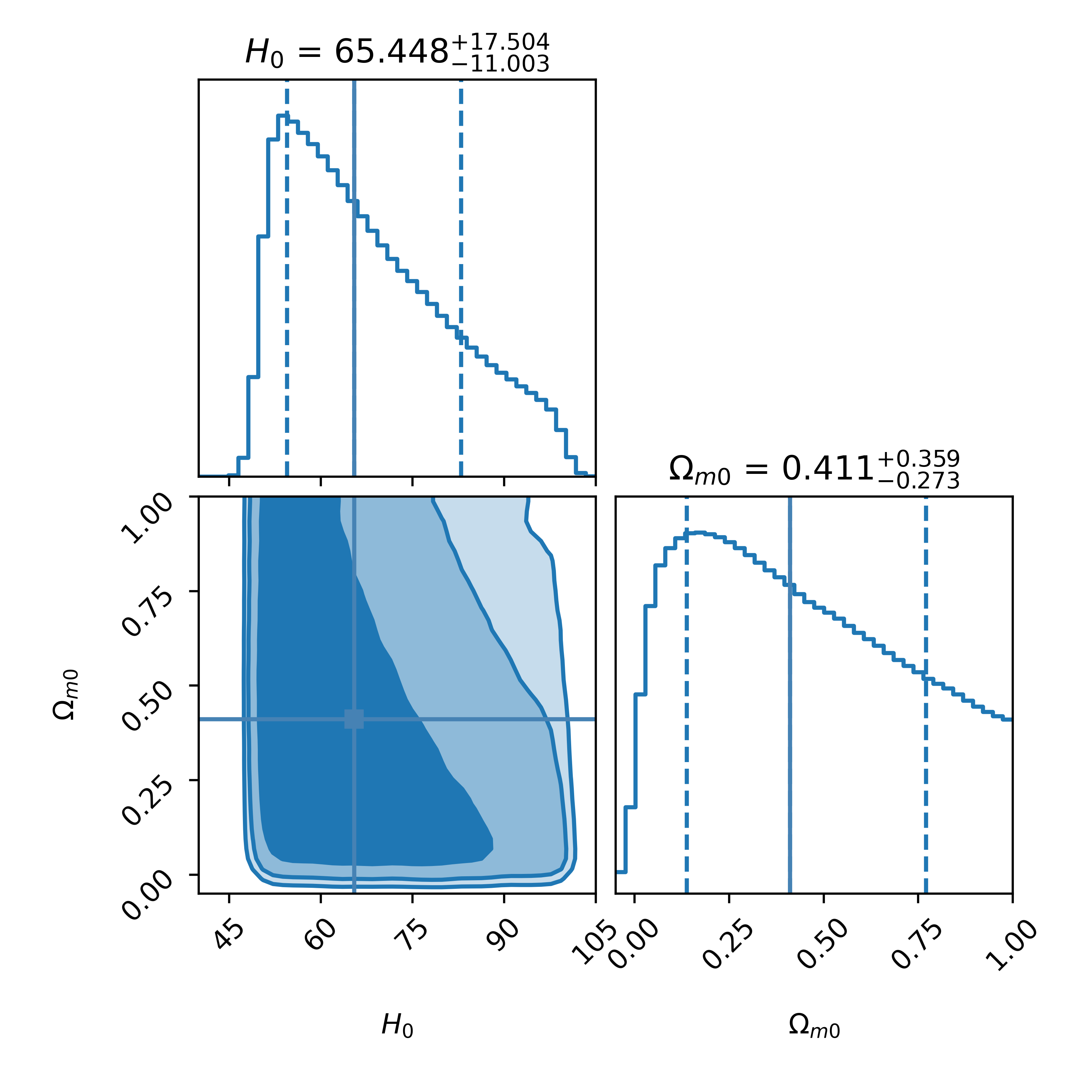}
\end{minipage}
\caption{The 1D and 2D posterior distributions of $H_0$ and $\Omega_{m0}$ obtained for $\Lambda$CDM model from GRB dataset with both constant $m~\&~c$ (\textbf{left figure)} and  dynamical $m~\&~c$ (\textbf{right figure}).}
    \label{csmlgy_lcdm_grbs}
\end{figure}

The 1D and 2D posterior distributions of $H_{0}$ and $\Omega_{m0}$ with both constant $m~\&~c$ and  dynamical $m~\&~c$ are shown in Figure \ref{csmlgy_lcdm_grbs}.

In both the cases with constant and dynamical Amati relation, bounds on $H_0$ and $\Omega_{m0}$ are not very strong as the error bars are wide. Also, as the two parameters are degenerate, a decrease in the best fit value of $H_0$ in the dynamical case relative to the constant one causes an  increase in the $\Omega_{m0}$ value.

\subsubsection*{2. \underline{ $\mathbf{q(z)}$ Parametrizations:}}
In this model, we consider three different parametrizations of the  deceleration parameter, $q(z)$. In each parametrization, we consider the constant Amati relation as well as the  dynamical Amati relation and put constraints on the coefficients of ${q(z)}$ parametrizations.\\

\noindent\textbf{P1:} $\mathbf{q(z)=q_0}$
\vspace{3mm}\\
Taking $q(z)$ to be a constant, the best fit values of $H_0$ and $q_0$ are given in Table \ref{tab_csmly_qcdm_p1_grbs}.
\begin{table}[H]
\begin{center}
\renewcommand{\arraystretch}{1.5}
    \begin{tabular}{ |c|c|c|c| } 
\hline
Parameters & Constant $m~\&~c$  & Dynamical $m~\&~c$ & Figure  \\  
\hline
 $H_{0}$ & $67.756^{+20.772}_{-20.557}$   & $67.985^{+18.802}_{-12.806}$ & \multirow{2}{*}{Figure \ref{csmlgy_qcdm_p1_grbs}} \\ 
 \cline{1-3}
 $q_{0}$ & $-0.277^{+0.666}_{-0.536}$     & $-0.317^{+0.702}_{-0.703}$  &  \\ 
\hline
\end{tabular}
\caption{The best fit values of $H_0$ and $q_{0}$ with 68\% confidence level obtained for P1 parametrization using GRB dataset.}
\label{tab_csmly_qcdm_p1_grbs}
\end{center}
\end{table}

\begin{figure}[H]
\centering
\begin{minipage}{.475\textwidth}
  \centering
  \includegraphics[width=1.0\linewidth]{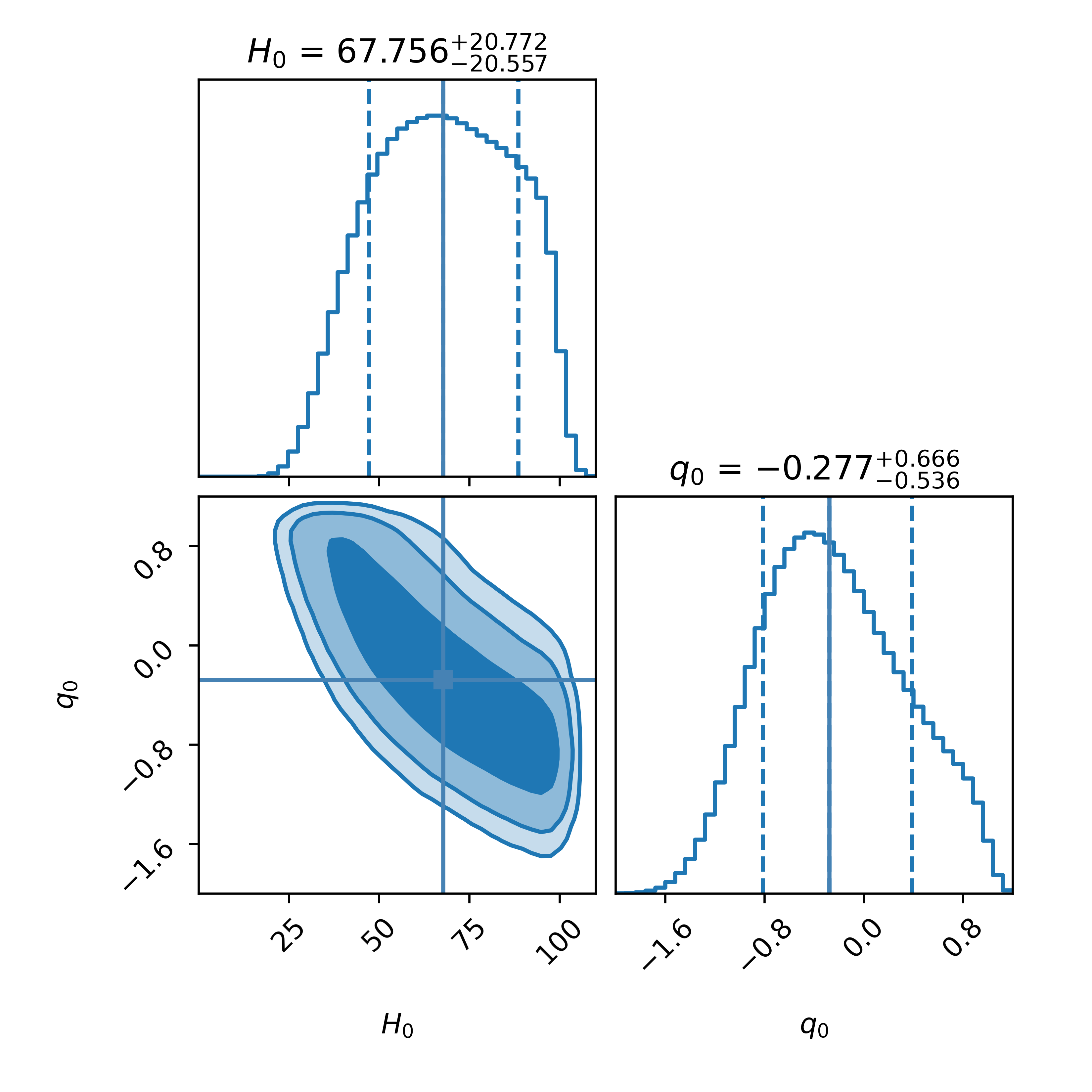}
\end{minipage}%
\hfill
\begin{minipage}{.475\textwidth}
  \centering
  \includegraphics[width=1.0\linewidth]{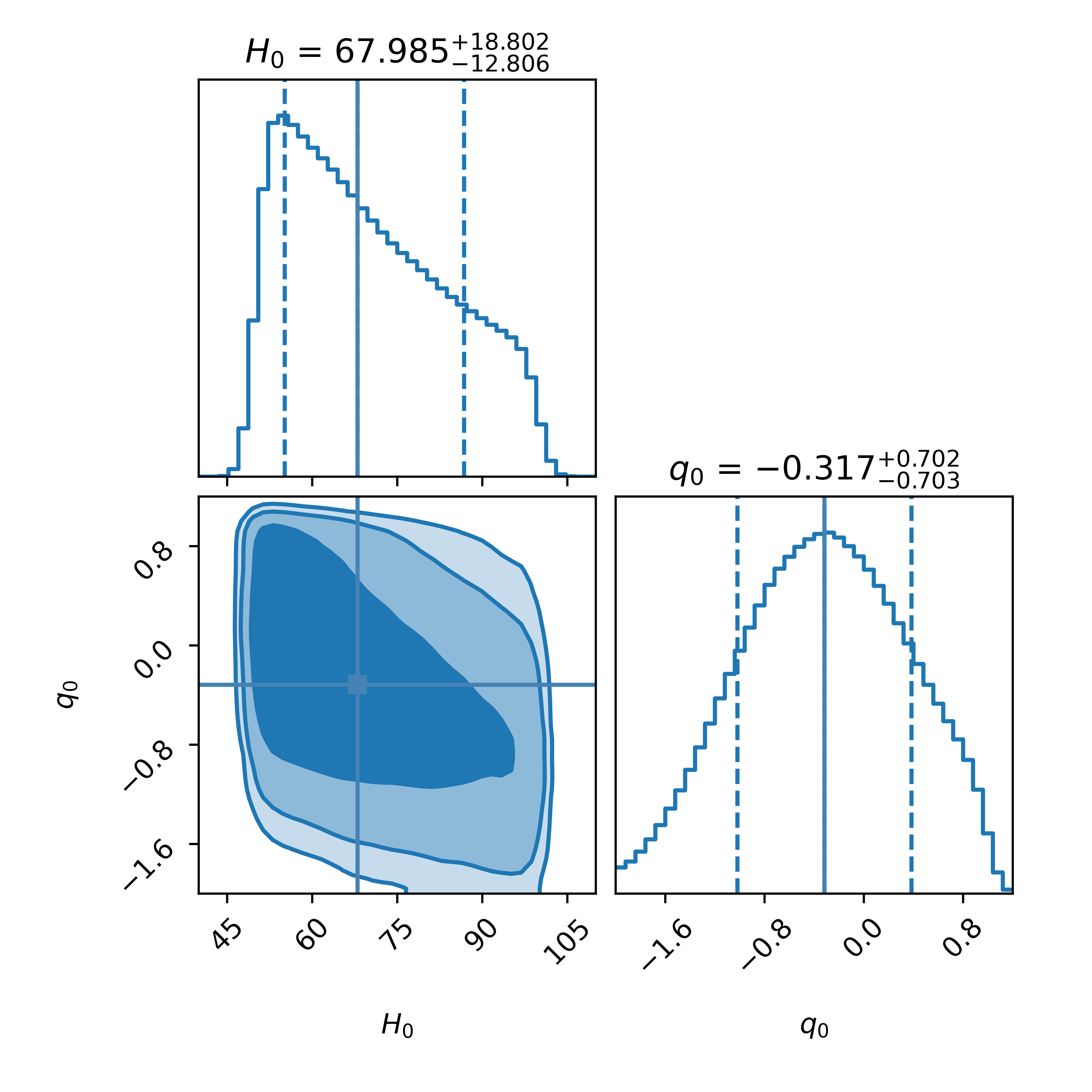}
\end{minipage}
\caption{The 1D and 2D posterior distributions of $H_0$ and $q_{0}$ obtained for P1 parametrization using GRB dataset with both constant $m~\&~c$ (\textbf{left figure)} and  dynamical $m~\&~c$ (\textbf{right figure}) .}
    \label{csmlgy_qcdm_p1_grbs}
\end{figure}

The 1D and 2D posterior distributions of $H_0$ and $q_{0}$ with 68\%, 95\% \& 99\% confidence levels for constant $m~\&~c$ and  dynamical $m~\&~c$ are shown in Figure \ref{csmlgy_qcdm_p1_grbs}.

\noindent\textbf{P2:} $\mathbf{q(z)=q_0+q_1z}$
\vspace{3mm}\\
In this parametrization, we consider $q(z)$ as a linear function of redshift. The best fit values of $H_0$, $q_0$ and $q_1$ are given in Table \ref{tab_csmly_qcdm_p2_grbs}.

\begin{figure}[H]
\centering
\begin{minipage}{.475\textwidth}
  \centering
  \includegraphics[width=1.0\linewidth]{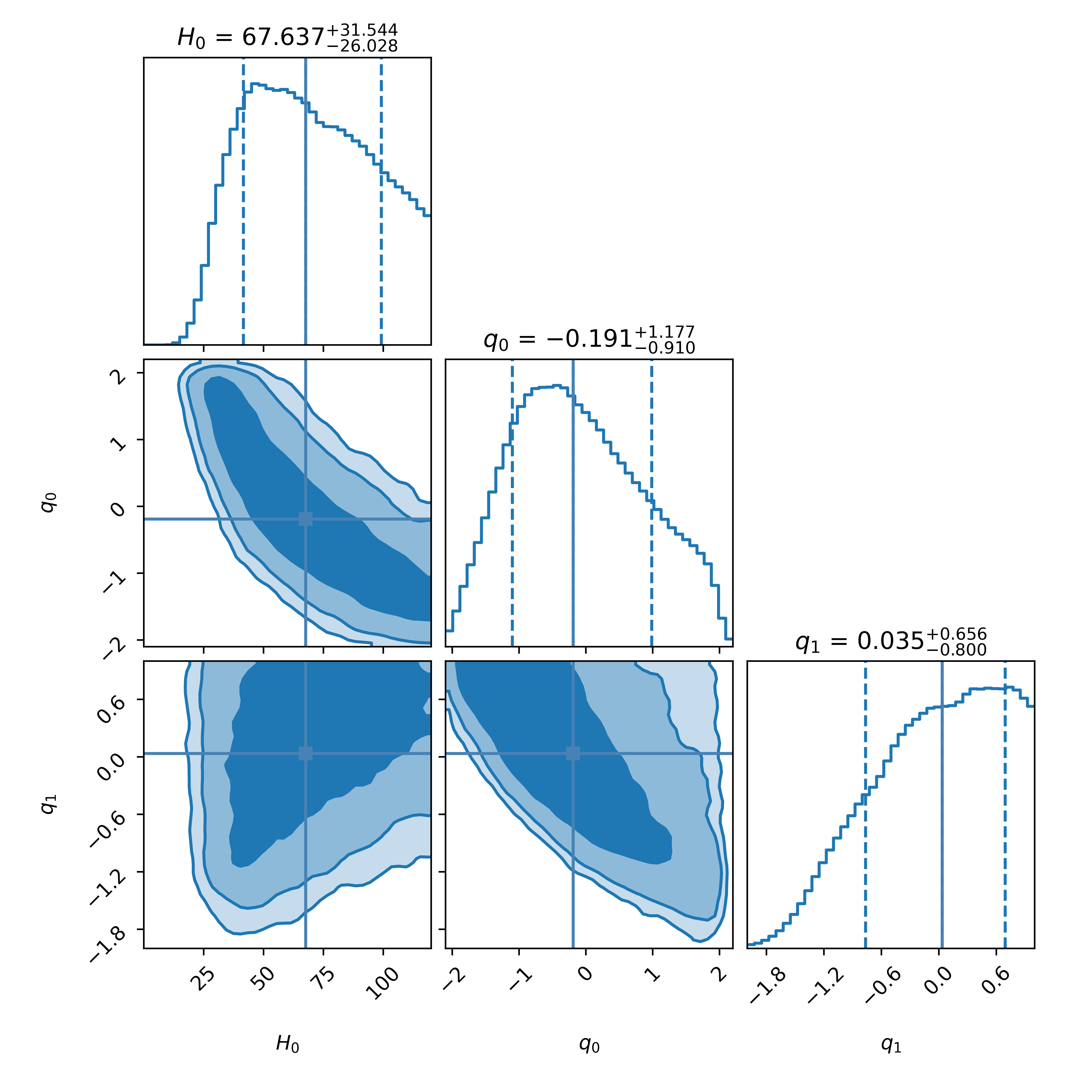}
\end{minipage}%
\hfill
\begin{minipage}{.475\textwidth} 
  \centering
  \includegraphics[width=1.0\linewidth]{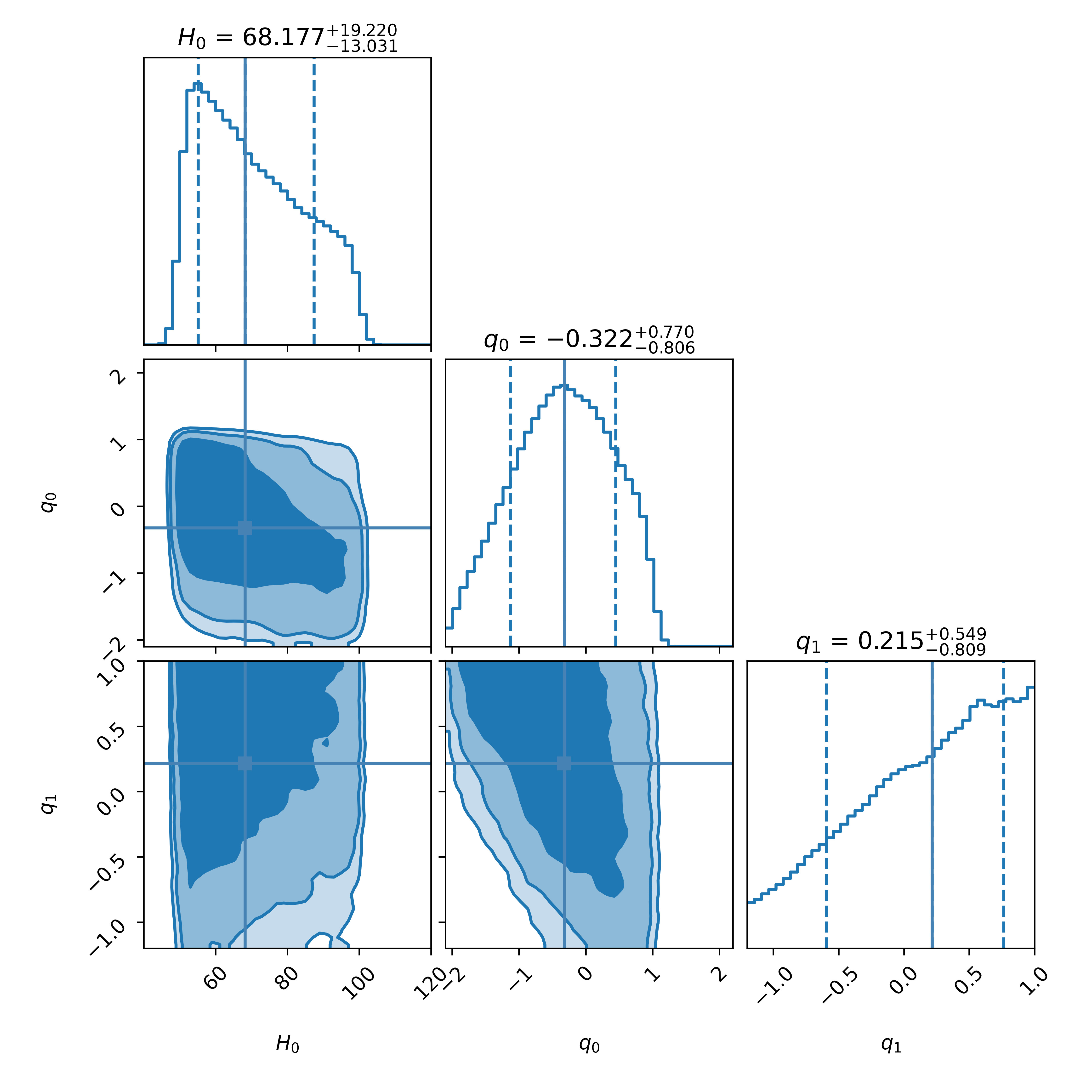}
\end{minipage}
\caption{The 1D and 2D posterior distributions of $H_0$, $q_{0}$ and $q_{1}$ obtained for P2 parametrization using GRB dataset with both constant $m~\&~c$ (\textbf{left figure)} and  dynamical $m~\&~c$ (\textbf{right figure}).}
    \label{csmlgy_qcdm_p2_grbs}
\end{figure}

\begin{table}[H]
\begin{center}
\renewcommand{\arraystretch}{1.5}
\begin{tabular}{ |c|c|c|c| } 
\hline
Parameters & Constant $m~\&~c$  & Dynamical $m~\&~c$  & Figure  \\
\hline
$H_0$ & $67.637^{+31.544}_{-26.028}$ & $68.177^{+19.220}_{-13.031}$ & \multirow{3}{*}{Figure \ref{csmlgy_qcdm_p2_grbs}} \\ 
\cline{1-3}
 $q_{0}$ & $-0.191^{+1.177}_{-0.910}$   &$-0.322^{+0.770}_{-0.806}$ & \\
\cline{1-3}
$q_{1}$ & $0.035^{+0.656}_{-0.800}$   &$0.215^{+0.549}_{-0.809}$ &  \\ 
\hline
\end{tabular}
\caption{The best fit values of $H_0$, $q_{0}$ and $q_{1}$ with 68\% confidence level obtained for P2 parametrization using GRB dataset.}
\label{tab_csmly_qcdm_p2_grbs}
\end{center}
\end{table}
The 1D and 2D posterior distributions of $H_0$, $q_{0}$ and $q_{1}$ with 68\%, 95\% \& 99\% confidence levels for both constant $m~\&~c$ and  dynamical $m~\&~c$ are shown in Figure \ref{csmlgy_qcdm_p2_grbs}.

\noindent\textbf{P3:} $\mathbf{q(z)=q_0+q_1\dfrac{z}{1+z}}$
\vspace{3mm}\\
In this parametrization, we choose $q(z)$ as a function of redshift which converges at high redshift. The best fit values of $H_0$, $q_0$ and $q_1$ are given in Table \ref{tab_csmly_qcdm_p3_grbs}. 

\begin{table}[H]
\begin{center}
\renewcommand{\arraystretch}{1.5} 
    \begin{tabular}{ |c|c|c|c| } 
\hline
Parameters & Constant $m~\&~c$  & Dynamical $m~\&~c$ & Figure   \\
\hline
 $H_{0}$ & $70.966^{+30.264}_{-27.42}$   &$67.679^{+18.834}_{-12.768}$ & \multirow{3}{*}{Figure \ref{csmlgy_qcdm_p3_grbs}} \\ 
\cline{1-3}
 $q_{0}$ & $-0.350^{+1.017}_{-0.790}$   &$-0.348^{+0.760}_{-0.787}$ & \\ 
\cline{1-3} 
$q_{1}$ & $0.201^{+1.227}_{-1.397}$   &$0.402^{+1.130}_{-1.465}$ &   \\ 
\hline
\end{tabular}
\caption{The best fit values of $H_0$, $q_{0}$ and $q_{1}$ with 68\% confidence level obtained for P3 parametrization using GRB dataset.}
\label{tab_csmly_qcdm_p3_grbs}
\end{center}
\end{table}

\begin{figure}[H]
\centering
\begin{minipage}{.475\textwidth}
  \centering
  \includegraphics[width=1.0\linewidth]{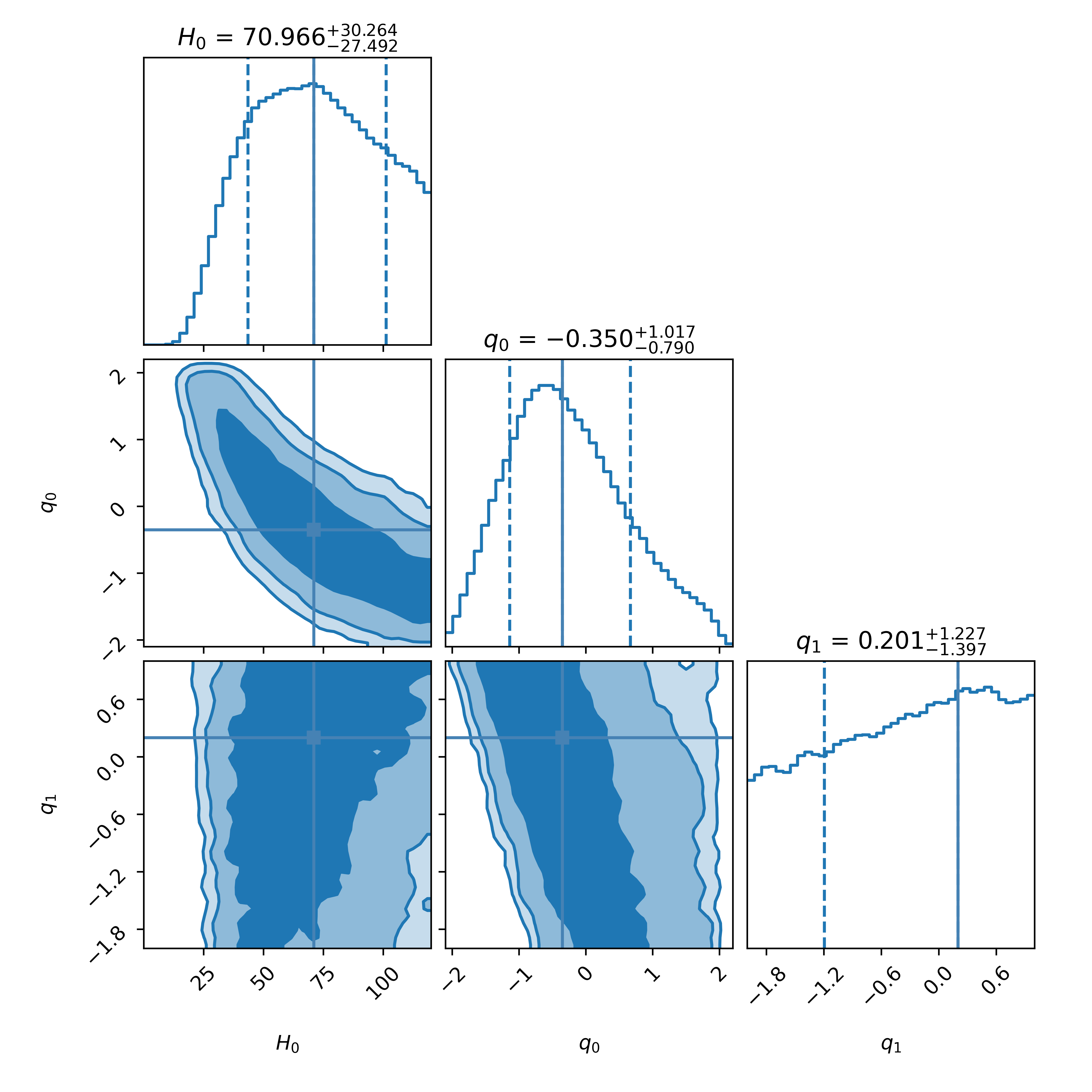}
\end{minipage}%
\hfill
\begin{minipage}{.475\textwidth}
  \centering
  \includegraphics[width=1.0\linewidth]{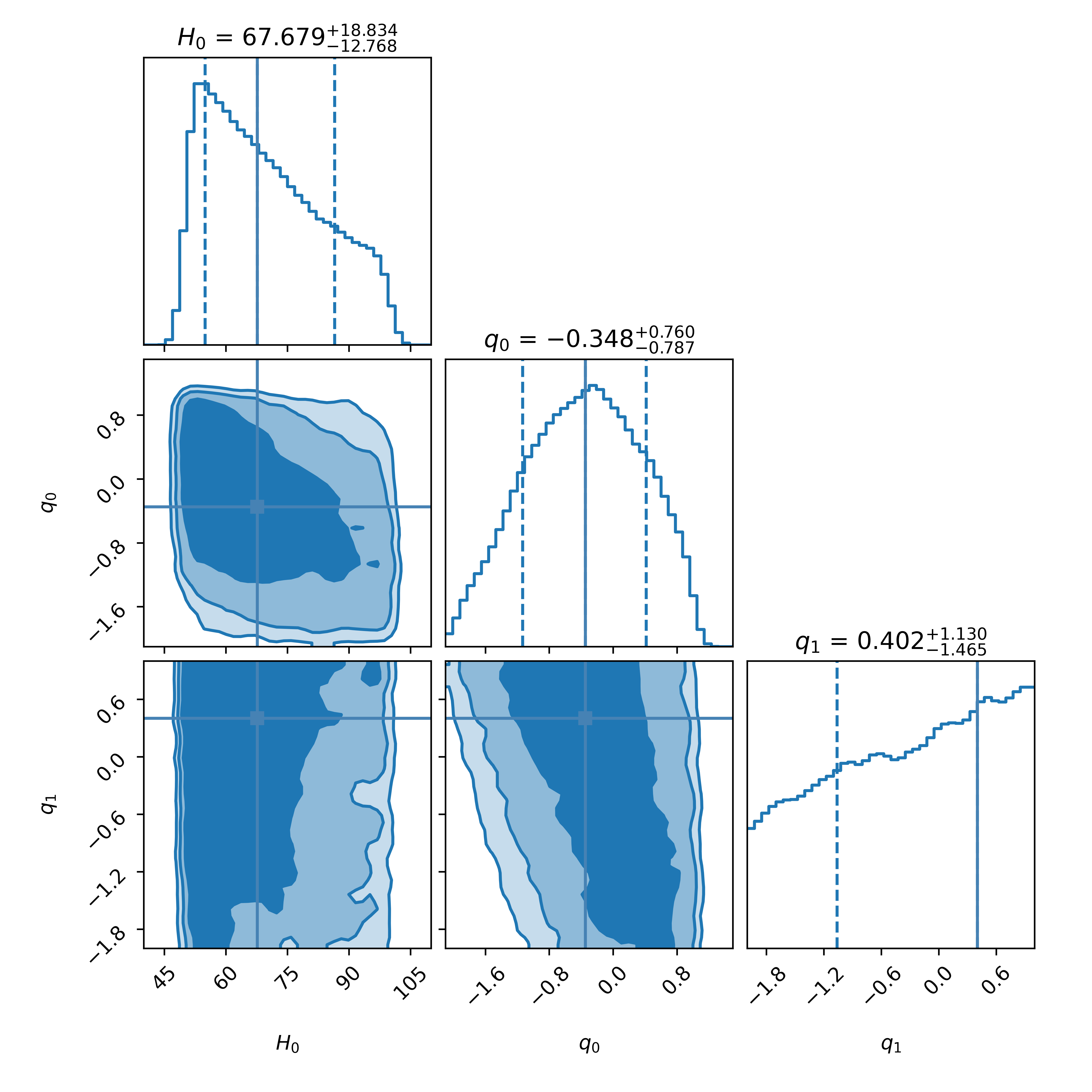}
\end{minipage}
\caption{The 1D and 2D posterior distributions of $H_0$, $q_{0}$ and $q_{1}$ obtained for P3 parametrization from GRB dataset with both constant $m~\&~c$ (\textbf{left figure)} and  dynamical $m~\&~c$ (\textbf{right figure}).}
    \label{csmlgy_qcdm_p3_grbs}
\end{figure}

The 1D and 2D posterior distributions of $H_0$, $q_{0}$ and $q_{1}$ with 68\%, 95\% \& 99\% confidence levels obtained from GRB dataset for both constant $m~\&~c$ and  dynamical $m~\&~c$ are shown in Figure \ref{csmlgy_qcdm_p3_grbs}.

In all parametrizations of $q(z)$, we find weak constraint on $q_0$. On the other hand, for parameter $q_1$, there is no constraint in both constant and dynamical Amati relations. Also, we find a negative value of $q_0$ and hence it suggests that the cosmic expansion is accelerating, which is the case in the current era of the universe. Here too, the constraint on $H_0$ is not very strong since the error bars are quite large.


\subsubsection*{3. \underline{ $\mathbf{z_t}$CDM Model:}}
In this model, we write $\Lambda$CDM model in terms of the  transition redshift $(z_t)$. The best fit values of $H_0$ and $z_t$ parameters are given in Table \ref{tab_csmly_ztcdm_grbs}.

\begin{table}[H]
\begin{center}
\renewcommand{\arraystretch}{1.5}
    \begin{tabular}{ |c|c|c|c| } 
\hline
Parameters & Constant $m~\&~c$  & Dynamical $m~\&~c$ & Figure   \\
\hline
 $H_{0}$ & $71.440^{+15.027}_{-12.698}$   &$67.925^{+17.488}_{-12.433}$ & \multirow{2}{*}{Figure \ref{csmlgy_ztcdm_grbs}} \\ 
 \cline{1-3}
 $z_{t}$ & $1.087^{+0.420}_{-0.486}$   &$1.047^{+0.637}_{-0.658}$ &  \\ 
\hline
\end{tabular}
\caption{The best fit values of $H_0$ and $z_{t}$ with 68\% confidence level obtained for $z_t$CDM model using GRB dataset.}
\label{tab_csmly_ztcdm_grbs}
\end{center}
\end{table}

\begin{figure}[H]
\centering
\begin{minipage}{.475\textwidth}
  \centering
  \includegraphics[width=1.0\linewidth]{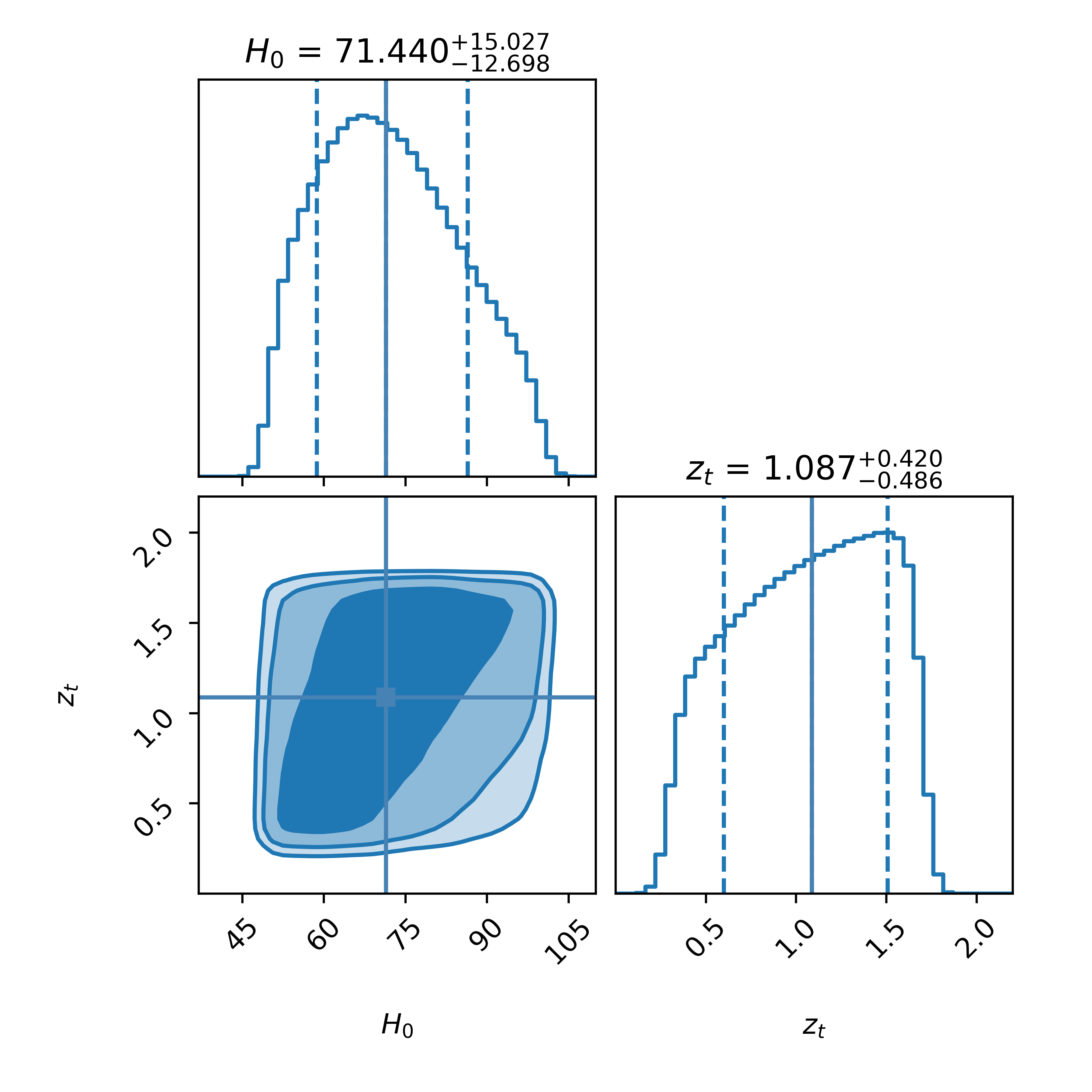}
\end{minipage}%
\hfill
\begin{minipage}{.475\textwidth}
  \centering
  \includegraphics[width=1.0\linewidth]{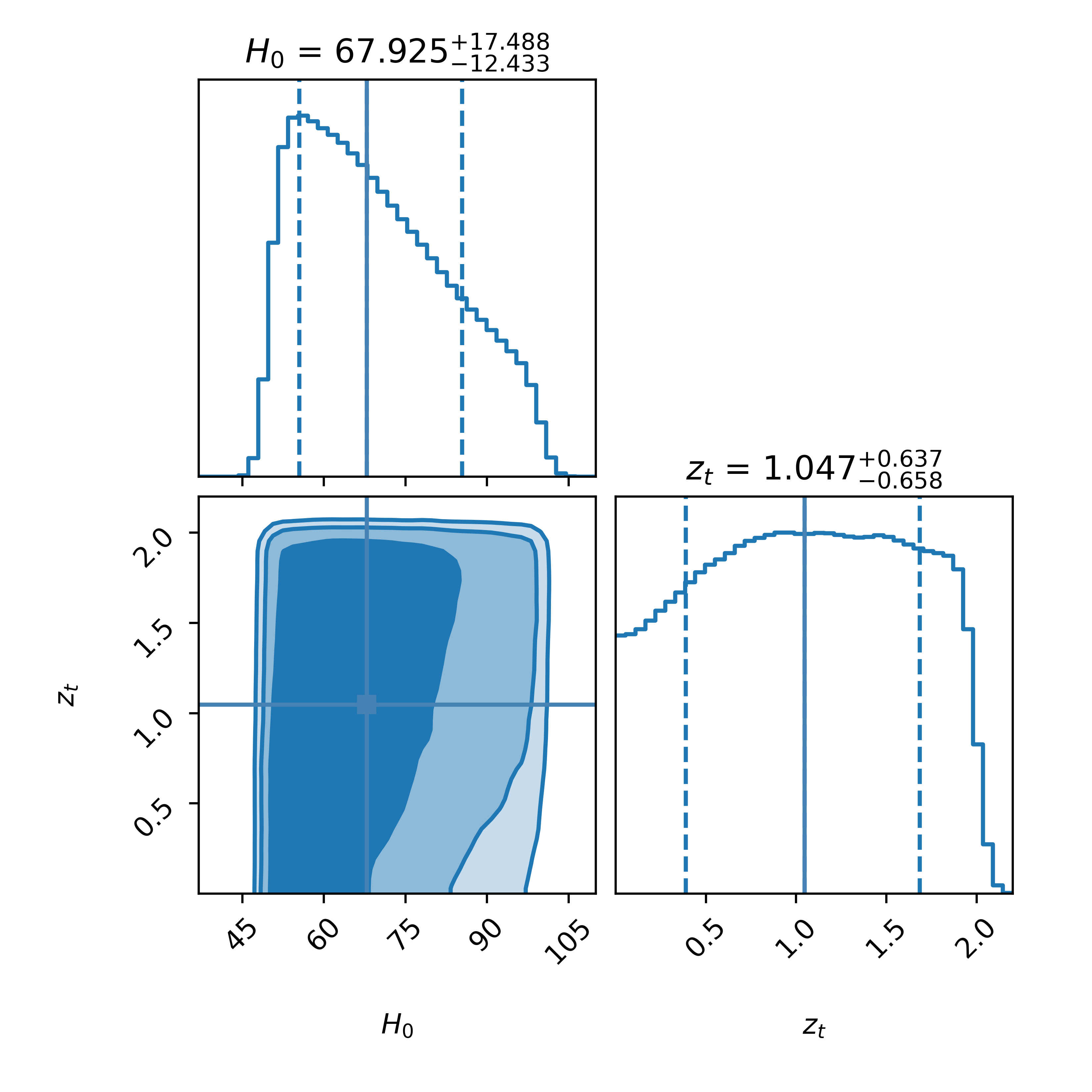}
\end{minipage}
\caption{The 1D and 2D posterior distributions of $H_0$ and $z_{t}$ for $z_t$CDM model with constants $m~\&~c$ (\textbf{left figure)} and  dynamical $m~\&~c$ (\textbf{right figure}).}
    \label{csmlgy_ztcdm_grbs}
\end{figure}

The 1D and 2D posterior distributions of $H_0$ and $z_{t}$ with 68\%, 95\% \& 99\% confidence levels obtained from GRB dataset for both constant $m~\&~c$ and  dynamical $m~\&~c$ are shown in Figure \ref{csmlgy_ztcdm_grbs}.

For both constant and dynamical Amati relations, we do not get any strong bound on $z_t$ with GRB data. 

\subsubsection{Joint fit with Gamma Rays Bursts and Type Ia SNe Datasets}

In order to study the effect of other cosmological datasets along with GRB, we repeat the above analysis by adding Pantheon data of Type Ia SNe to GRB dataset. We treat $M_B$ as a free parameter.


\subsubsection*{1. \underline{ $\Lambda$CDM Model:}}
The best fit values of $H_0$, $\Omega_{m0}$  and $M_B$ obtained for this model are given in Table \ref{tab_csmly_lcdm_grbs_snia}.

\begin{table}[H]
\begin{center}
\renewcommand{\arraystretch}{1.5}
    \begin{tabular}{ |c|c|c|c| } 
\hline
Parameters & Constant $m~\&~c$  & Dynamical $m~\&~c$ & Figure   \\
\hline
$H_0$ & $64.802^{+13.918}_{-11.412}$ & $66.248^{+15.133}_{-15.246}$ & \multirow{3}{*}{Figure \ref{csmlgy_lcdm_grbs_snia}} \\ 
\cline{1-3} 
 $\Omega_{m0}$ & $0.284^{+0.009}_{-0.009}$   &$0.285^{+0.009}_{-0.009}$ &  \\ 
\cline{1-3}
 $M_{B}$ & $-19.524^{+0.422}_{-0.420}$   &$-19.476^{+0.446}_{-0.568}$ &  \\ 
\hline
\end{tabular}
\caption{The best fit values of $H_0$, $\Omega_{m0}$  and $M_B$ with 68\% confidence level obtained for $\Lambda$CDM model using GRBs+Type Ia SNe datasets.}
\label{tab_csmly_lcdm_grbs_snia}
\end{center}
\end{table}

\begin{figure}[H]
\centering
\begin{minipage}{.475\textwidth}
  \centering
  \includegraphics[width=1.0\linewidth]{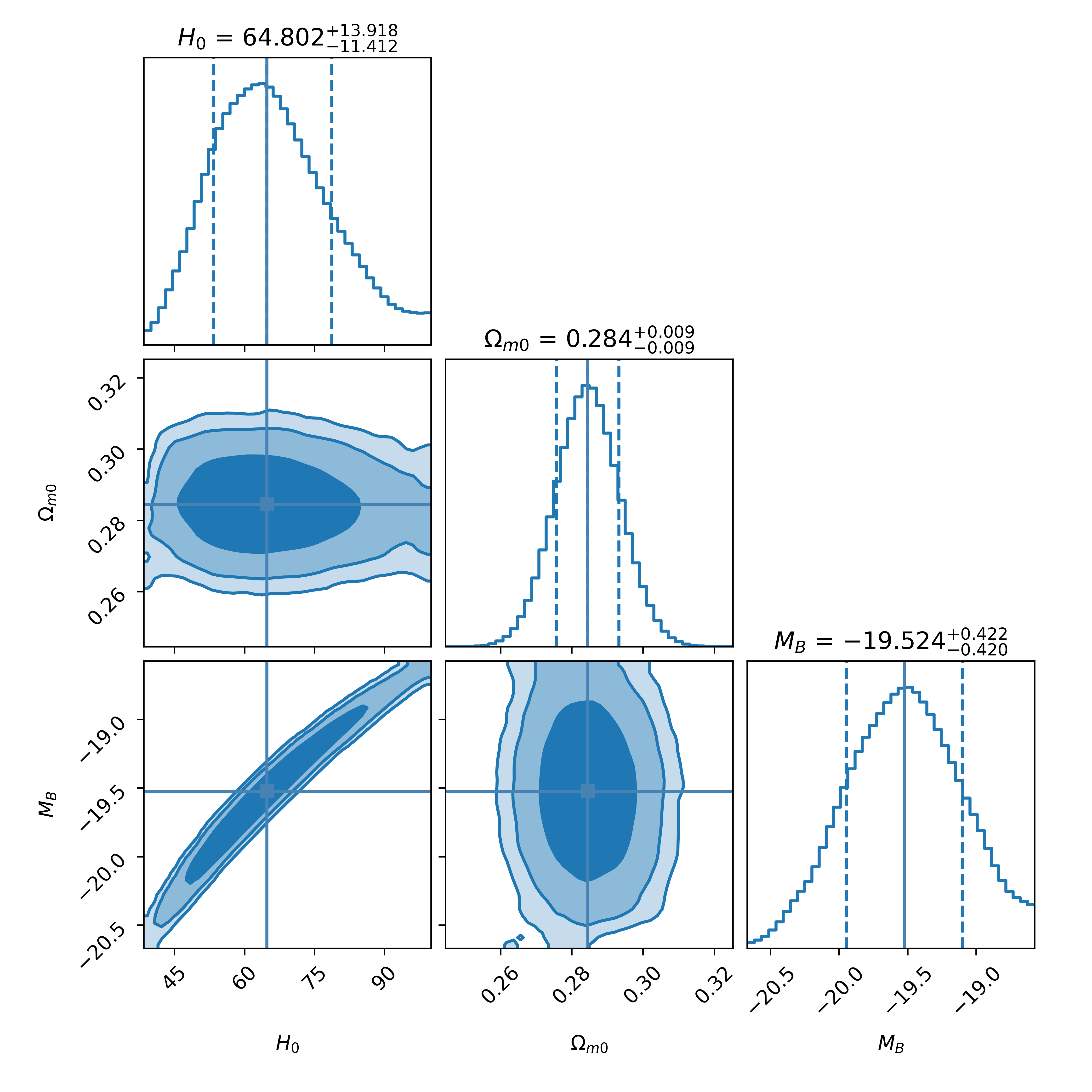}
\end{minipage}%
\hfill
\begin{minipage}{.475\textwidth}
  \centering
  \includegraphics[width=1.0\linewidth]{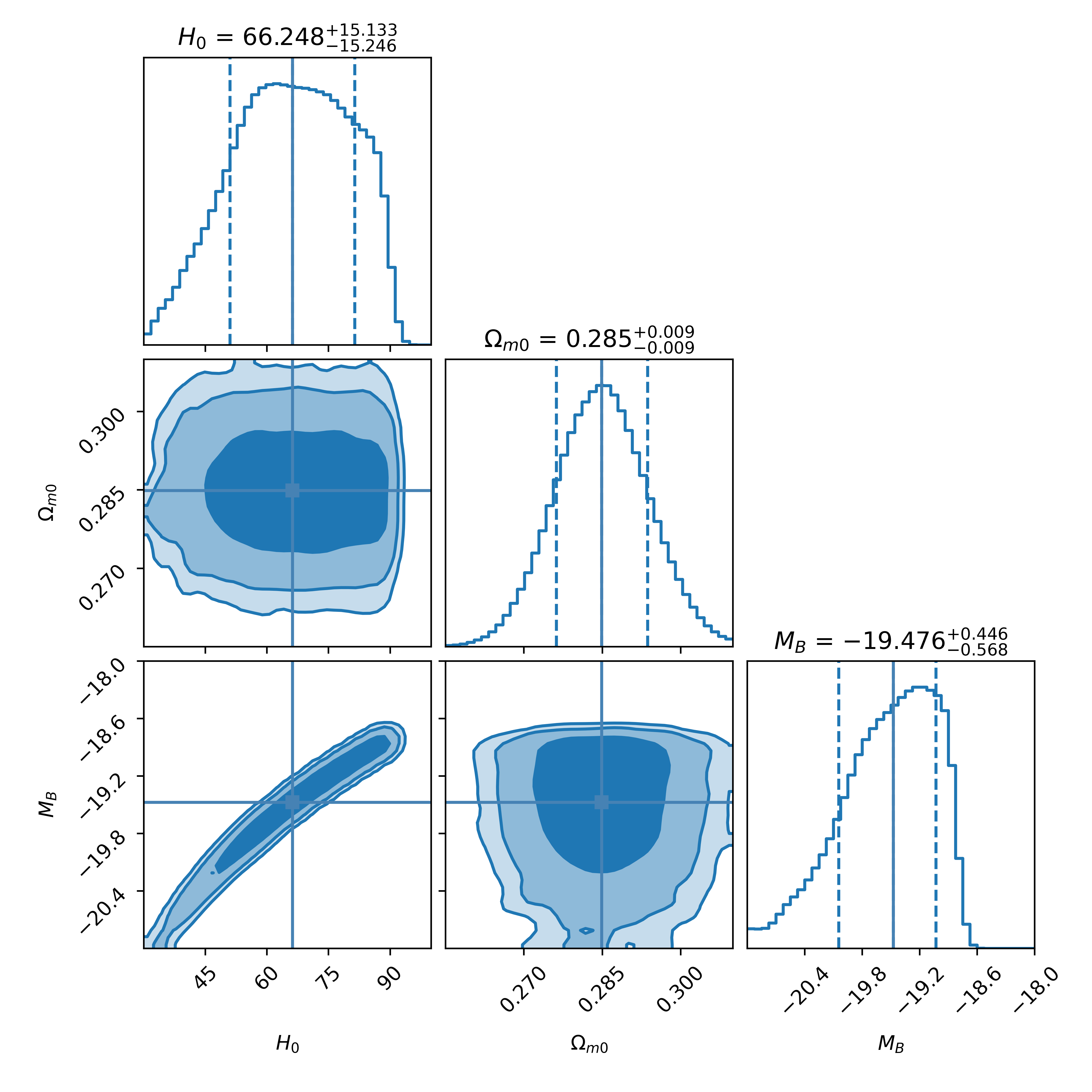}
\end{minipage}
\caption{The 1D and 2D posterior distributions of $H_0$, $z_{t}$ and $M_B$ obtained for $\Lambda$CDM model from GRBs+Type Ia SNe datasets for both constant $m~\&~c$ (\textbf{left figure)} and  dynamical $m~\&~c$ (\textbf{right figure}).}
    \label{csmlgy_lcdm_grbs_snia}
\end{figure}

The 1D and 2D posterior distributions of $H_0$, $\Omega_{m0}$  and $M_B$ with 68\%, 95\% \& 99\% confidence levels for both constant $m~\&~c$ and  dynamical $m~\&~c$ are shown in Figure \ref{csmlgy_lcdm_grbs_snia}.

\subsubsection*{2. \underline{ $\mathbf{q(z)}$ Parametrizations:}}

\noindent\textbf{P1:} $q(z)=q_0$
\vspace{3mm}\\
In first parametrization of $q(z)$, we consider it to be constant. The best fit value of $H_0$, $q_0$ and $M_B$ are shown in Table \ref{tab_csmly_qcdm_p1_grbs_snia}.

\begin{table}[H]
\begin{center} 
\renewcommand{\arraystretch}{1.5}
    \begin{tabular}{ |c|c|c|c| } 
\hline
Parameters & Constant $m~\&~c$  & Dynamical $m~\&~c$  & Figure  \\
\hline
$H_0$ & $70.236^{+13.698}_{-12.172}$ & $69.660^{+17.839}_{-16.716}$ & \multirow{3}{*}{Figure \ref{csmlgy_qcdm_p1_grbs_snia}} \\ 
\cline{1-3} 
 $q_{0}$ & $-0.342^{+0.015}_{-0.015}$   &$-0.342^{+0.015}_{-0.015}$ &  \\ 
 \cline{1-3}
 $M_{B}$ & $-19.322^{+0.387}_{-0.414}$   & $-19.339^{+0.495}_{-0.596}$ &  \\ 
\hline
\end{tabular}
\caption{The best fit values of $H_{0}$, $q_{0}$  and $M_B$ with 68\% confidence level obtained  for P1 parametrization using GRBs+Type Ia SNe datasets.}
\label{tab_csmly_qcdm_p1_grbs_snia}
\end{center}
\end{table}

\begin{figure}[H]
\centering
\begin{minipage}{.475\textwidth}
  \centering
  \includegraphics[width=1.0\linewidth]{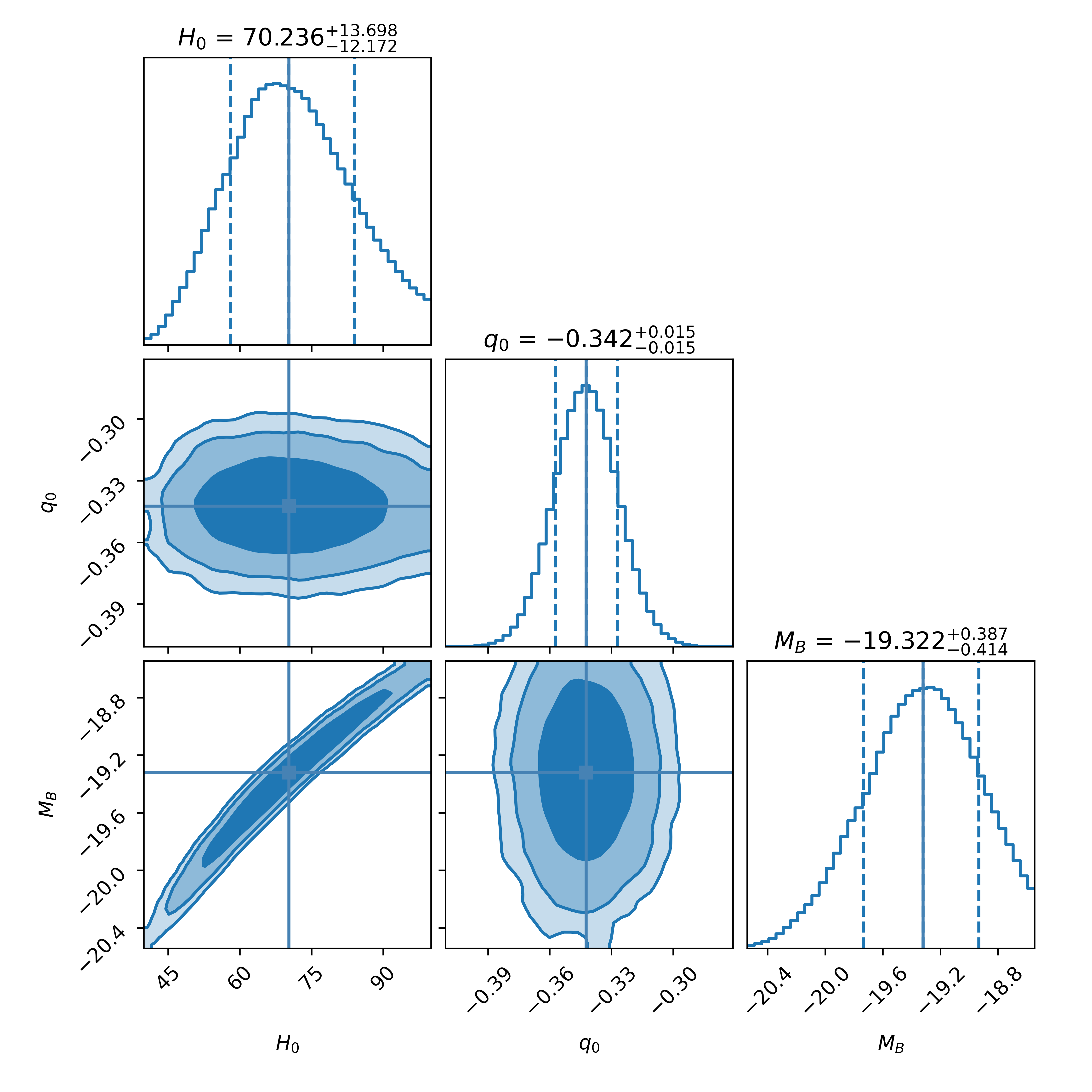}
\end{minipage}%
\hfill
\begin{minipage}{.475\textwidth}
  \centering
  \includegraphics[width=1.0\linewidth]{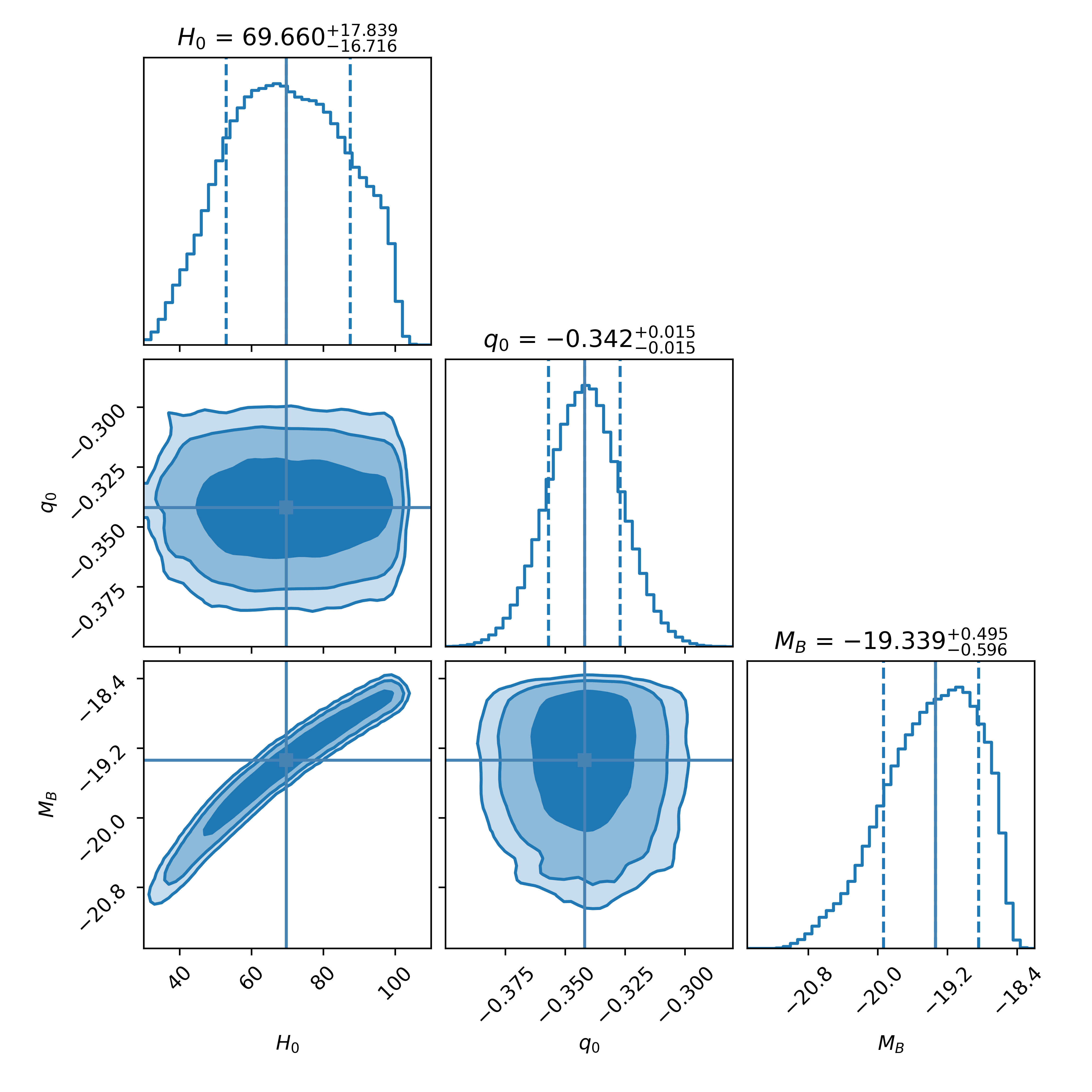}
\end{minipage}
\caption{The 1D and 2D posterior distributions of $H_0$, $q_{0}$ and $M_B$ obtained for P1 parametrization from GRBs+Type Ia SNe datasets with both constant $m~\&~c$ (\textbf{left figure)} and  dynamical $m~\&~c$ (\textbf{right figure}).} 
    \label{csmlgy_qcdm_p1_grbs_snia}
\end{figure}

The 1D and 2D posterior distributions of $H_{0}$, $q_{0}$ and $M_B$ with both constant $m~\&~c$ and dynamical $m~\&~c$ are shown in Figure \ref{csmlgy_qcdm_p1_grbs_snia}.

\noindent\textbf{P2:} $\mathbf{q(z)=q_0+q_1z}$
\vspace{3mm}\\
Deceleration parameter, $q(z)$, is regarded as a linear function of redshift in this parametrization. The best fit value of $H_0$, $q_0$, $q_1$ and $M_B$ are shown in Table \ref{tab_csmly_qcdm_p2_grbs_snia}. 
\begin{table}[H]
\begin{center}
\renewcommand{\arraystretch}{1.5}
    \begin{tabular}{ |c|c|c|c| } 
\hline
Parameters & Constant $m~\&~c$  & Dynamical $m~\&~c$ & Figure  \\
\hline
$H_0$ & $58.882^{+10.737}_{-08.250}$ & $69.596^{+17.127}_{-17.597}$ & \multirow{4}{*}{Figure \ref{csmlgy_qcdm_p2_grbs_snia}} \\ 
\cline{1-3} 
 $q_{0}$ & $-0.645^{+0.042}_{-0.042}$   &$-0.638^{+0.046}_{-0.042}$ &  \\ 
\cline{1-3} 
$q_{1}$ & $1.147^{+0.149}_{-0.148}$   &$1.133^{+0.148}_{-0.165}$ & \\ 
\cline{1-3}
 $M_{B}$ & $-19.741^{+0.362}_{-0.329}$   & $-19.380^{+0.479}_{-0.632}$ &  \\ 
\hline
\end{tabular}
\caption{The best fit values of $H_{0}$, $q_{0}$, $q_{1}$  and $M_B$ with 68\% confidence level obtained  for P2 parametrization using GRBs+Type Ia SNe datasets.}
\label{tab_csmly_qcdm_p2_grbs_snia}
\end{center}
\end{table}

\begin{figure}[H]
\centering
\begin{minipage}{.475\textwidth}
  \centering
  \includegraphics[width=1.0\linewidth]{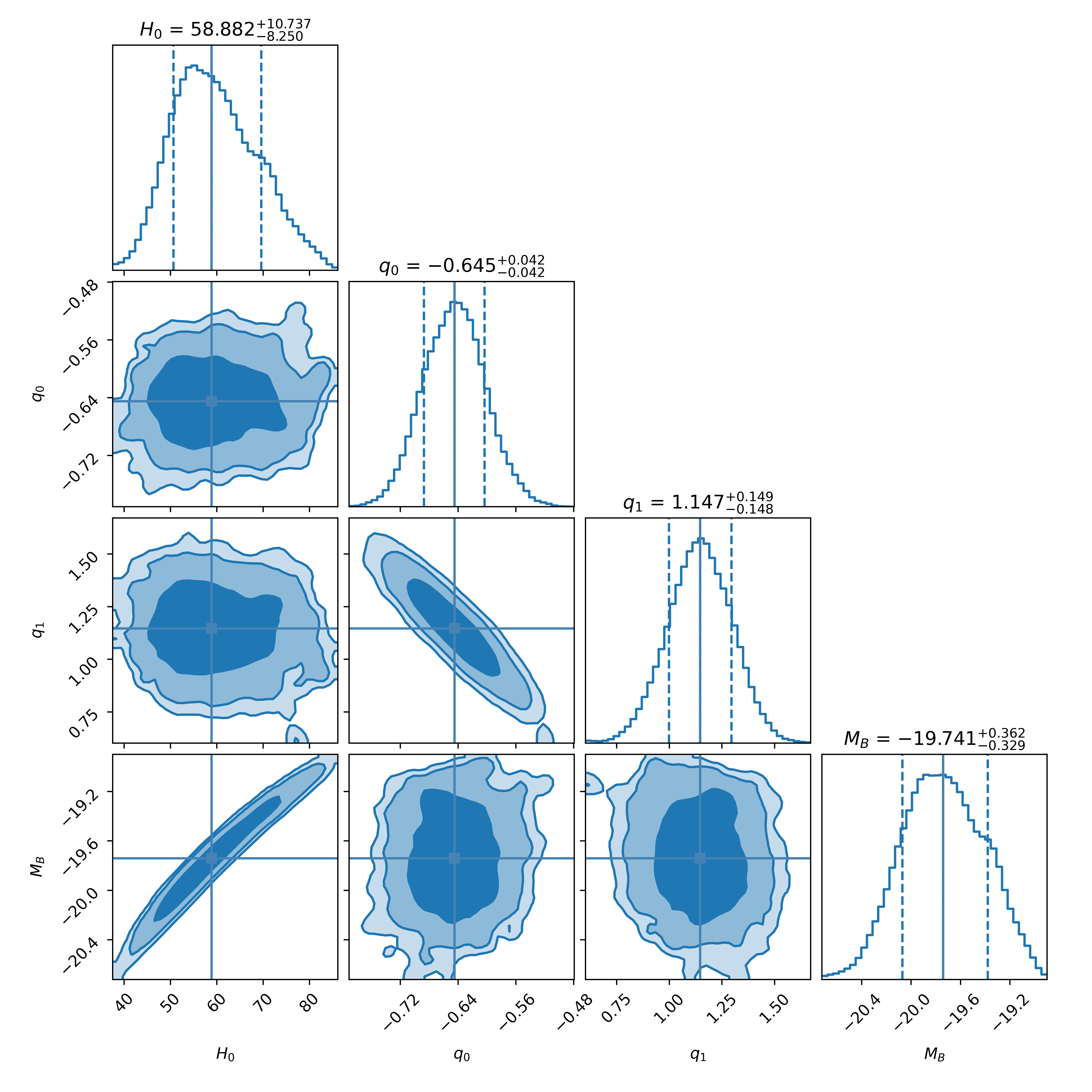}
\end{minipage}%
\hfill
\begin{minipage}{.475\textwidth}
  \centering
  \includegraphics[width=1.0\linewidth]{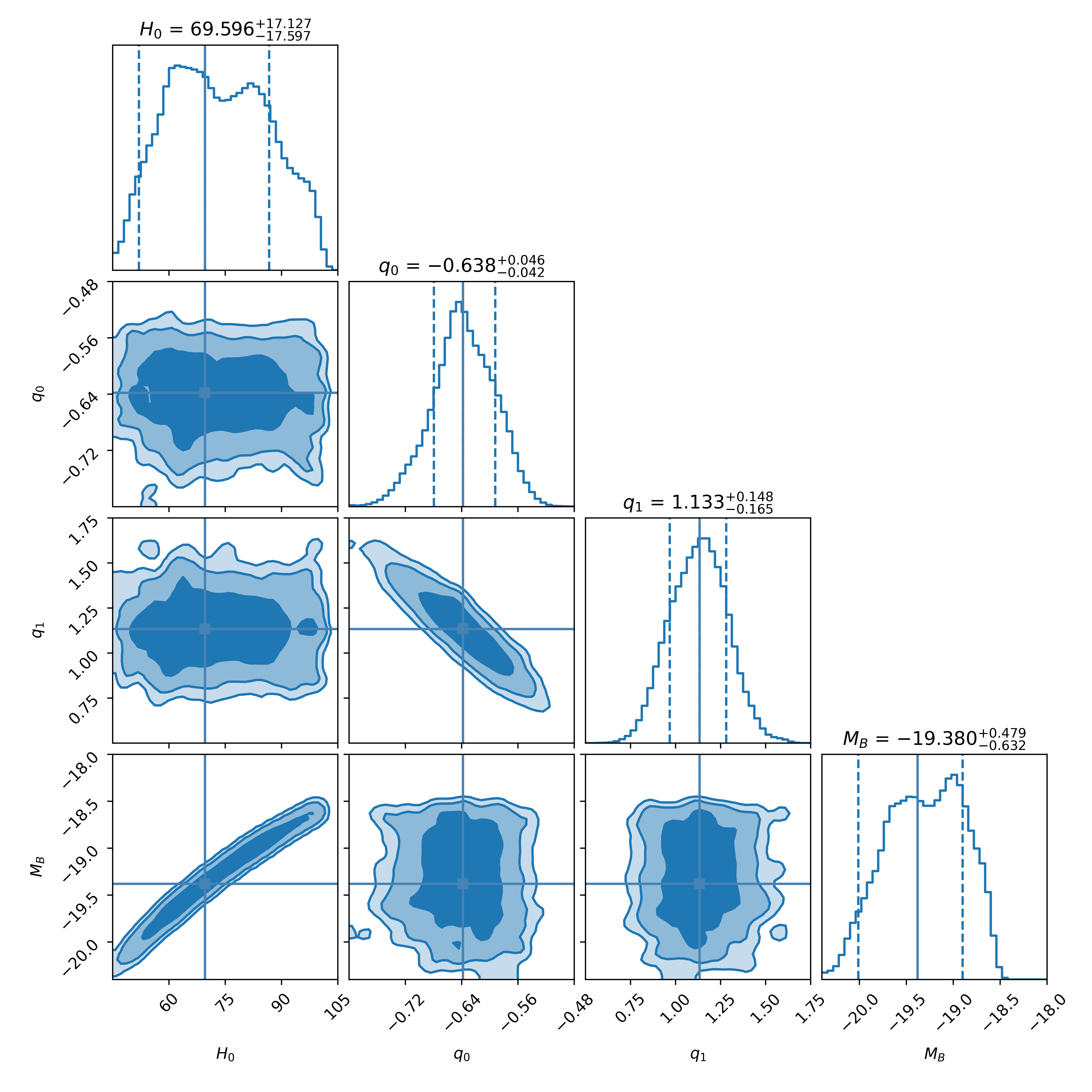}
\end{minipage}
\caption{The 1D and 2D posterior distributions of $H_{0}$, $q_{0}$, $q_{1}$ and $M_B$ obtained for P2 parametrization from GRBs+Type Ia SNe datasets with both constant $m~\&~c$ (\textbf{left figure)} and  dynamical $m~\&~c$ (\textbf{right figure}).}
    \label{csmlgy_qcdm_p2_grbs_snia}
\end{figure}

The 1D and 2D posterior distributions of $H_{0}$, $q_{0}$, $q_{1}$ and $M_B$ with both constant $m~\&~c$ and  dynamical $m~\&~c$ are shown in Figure \ref{csmlgy_qcdm_p2_grbs_snia}.

\noindent\textbf{P3:} $\mathbf{q(z)=q_0+q_1\dfrac{z}{1+z}}$
\vspace{3mm}\\
In this parameterization, we consider $q(z)$ as a redshift function that converges at high z. The best fit value of $H_0$, $q_0$, $q_1$ and $M_B$ are shown in Table \ref{tab_csmly_qcdm_p3_grbs_snia}. 

\begin{table}[H]
\begin{center}
\renewcommand{\arraystretch}{1.5}
    \begin{tabular}{ |c|c|c|c| } 
\hline
Parameters & Constant $m~\&~c$  & Dynamical $m~\&~c$ & Figure  \\
\hline
$H_0$ & $66.298^{+15.218}_{-10.710}$ & $70.189^{+16.885}_{-13.459}$ & \multirow{4}{*}{Figure \ref{csmlgy_qcdm_p3_grbs_snia}} \\ 
\cline{1-3}  
 $q_{0}$ & $-0.732^{+0.048}_{-0.049}$   &$-0.738^{+0.049}_{-0.049}$ & \\ 
\cline{1-3} 
$q_{1}$ & $2.039^{+0.243}_{-0.246}$   &$2.064^{+0.251}_{-0.245}$ & \\ 
\cline{1-3}
 $M_{B}$ & $-19.489^{+0.446}_{-0.384}$   & $-19.369^{+0.467}_{-0.461}$ &  \\ 
\hline
\end{tabular}
\caption{The best fit values of $H_{0}$, $q_{0}$, $q_{1}$ and $M_B$ with 68\% confidence level obtained  for P3 parametrization using GRBs+Type Ia SNe datasets.}
\label{tab_csmly_qcdm_p3_grbs_snia}
\end{center}
\end{table}

\begin{figure}[H]
\centering
\begin{minipage}{.475\textwidth}
  \centering
  \includegraphics[width=1.0\linewidth]{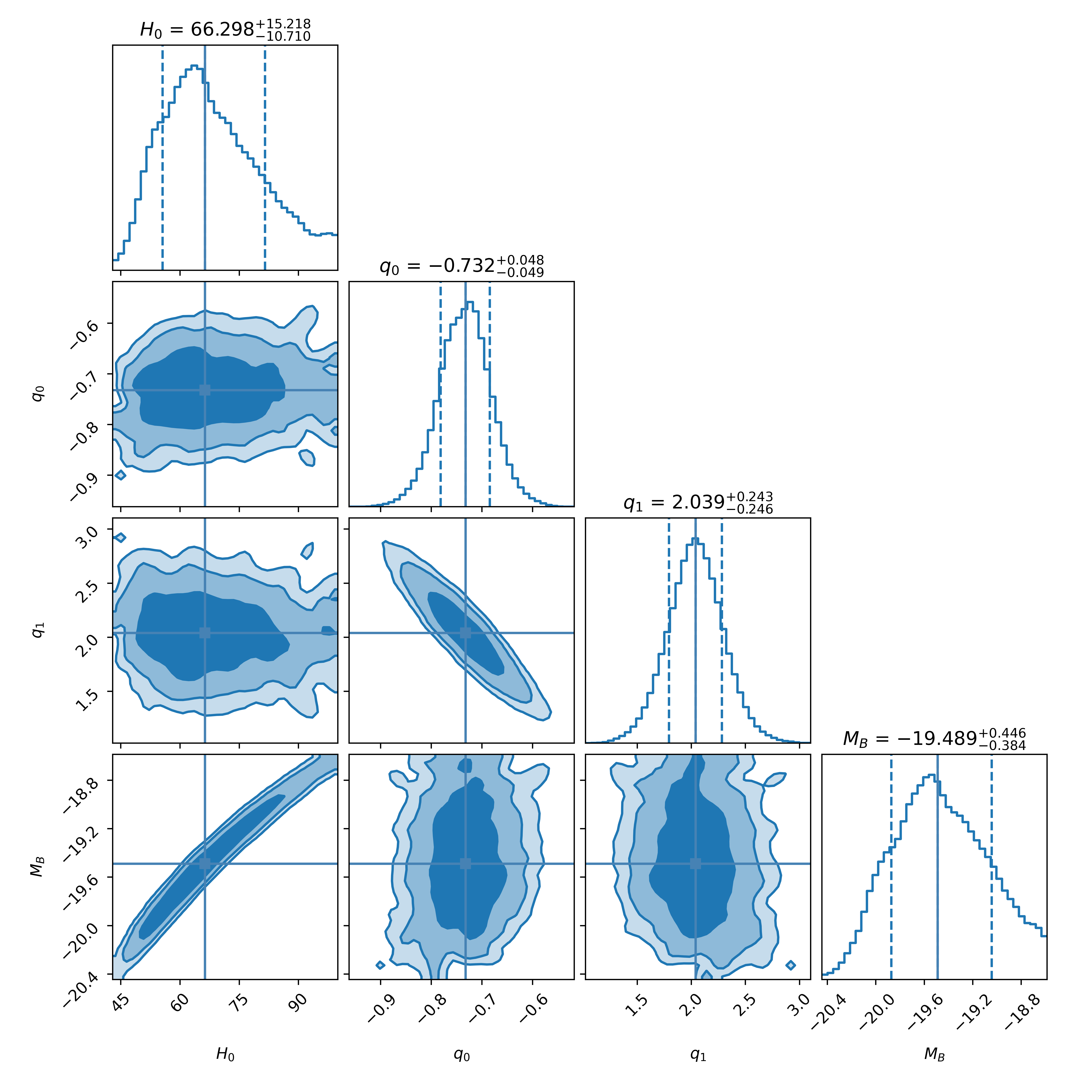}
\end{minipage}%
\hfill
\begin{minipage}{.475\textwidth}
  \centering
  \includegraphics[width=1.0\linewidth]{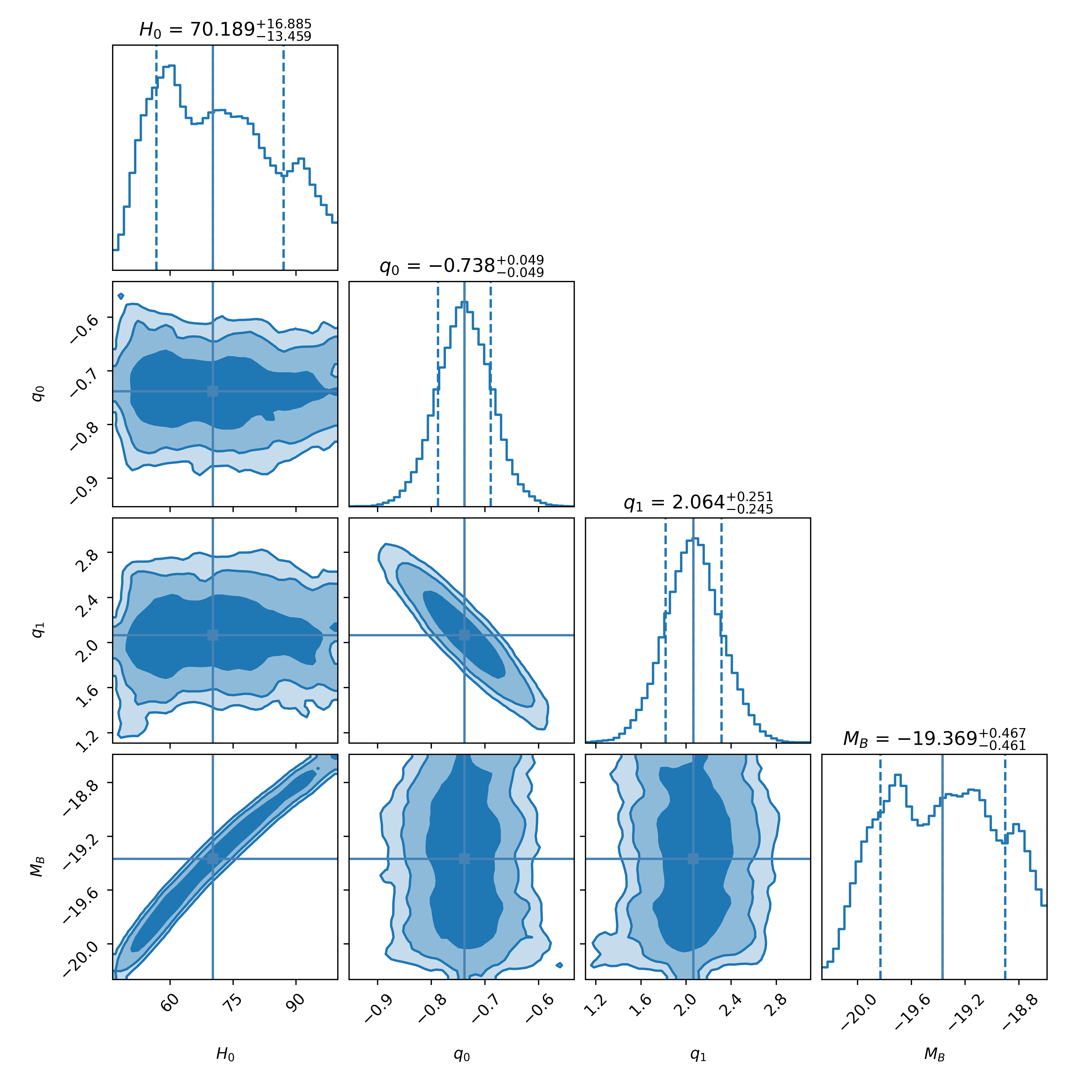}
\end{minipage}
\caption{The 1D and 2D posterior distributions of $H_{0}$, $q_{0}$, $q_{1}$ and $M_B$ obtained  for P3 parametrization from GRBs+Type Ia SNe datasets with both constant $m~\&~c$ (\textbf{left figure)} and  dynamical $m~\&~c$ (\textbf{right figure}).}
    \label{csmlgy_qcdm_p3_grbs_snia}
\end{figure}

The 1D and 2D posterior distributions of $H_{0}$, $q_{0}$, $q_{1}$ and $M_B$ with both constant $m~\&~c$ and  dynamical $m~\&~c$ are shown in Figure \ref{csmlgy_qcdm_p3_grbs_snia}.

For all parametrizations of $q(z)$, we find a very tight constraint on $q_0$ and $q_1$ for both constant and dynamical Amati relations. Furthermore, our results indicate a negative value of $q_0$, implying that the cosmic expansion is currently accelerating.


\subsubsection*{3. \underline{ $\mathbf{z_t}$CDM Model:}}
In this model, we write $\Lambda$CDM model in terms of the transition redshift $(z_t)$. The best fit values of $H_0$, $z_t$ and $M_B$ are given in Table \ref{tab_csmly_ztcdm_grbs_snia}.

\begin{table}[H]
\begin{center}
\renewcommand{\arraystretch}{1.5}
    \begin{tabular}{ |c|c|c|c| } 
\hline
Parameters & Constant $m~\&~c$  & Dynamical $m~\&~c$ & Figure   \\
\hline
$H_0$ & $65.380^{+13.072}_{-9.433}$ & $62.868^{+21.376}_{-16.514}$ & \multirow{2}{*}{Figure \ref{csmly_w_GRB_SNIa_ztCDM_fig}} \\ 
\cline{1-3}
 $z_{t}$ & $0.713^{+0.026}_{-0.025}$   &$0.713^{+0.025}_{-0.025}$ &  \\
 \cline{1-3}
 $M_{B}$ & $-19.505^{+0.396}_{-0.337}$   & $-19.590^{+0.637}_{-0.661}$ &  \\ 
\hline
\end{tabular} 
\caption{The best fit values of $H_0$, $z_{t}$ and $M_B$ with 68\% confidence level obtained  for $z_t$CDM model using GRBs+Type Ia SNe datasets.}
\label{tab_csmly_ztcdm_grbs_snia}
\end{center}
\end{table}  

\begin{figure}[H]
\centering
\begin{minipage}{.475\textwidth}
  \centering
  \includegraphics[width=1.0\linewidth]{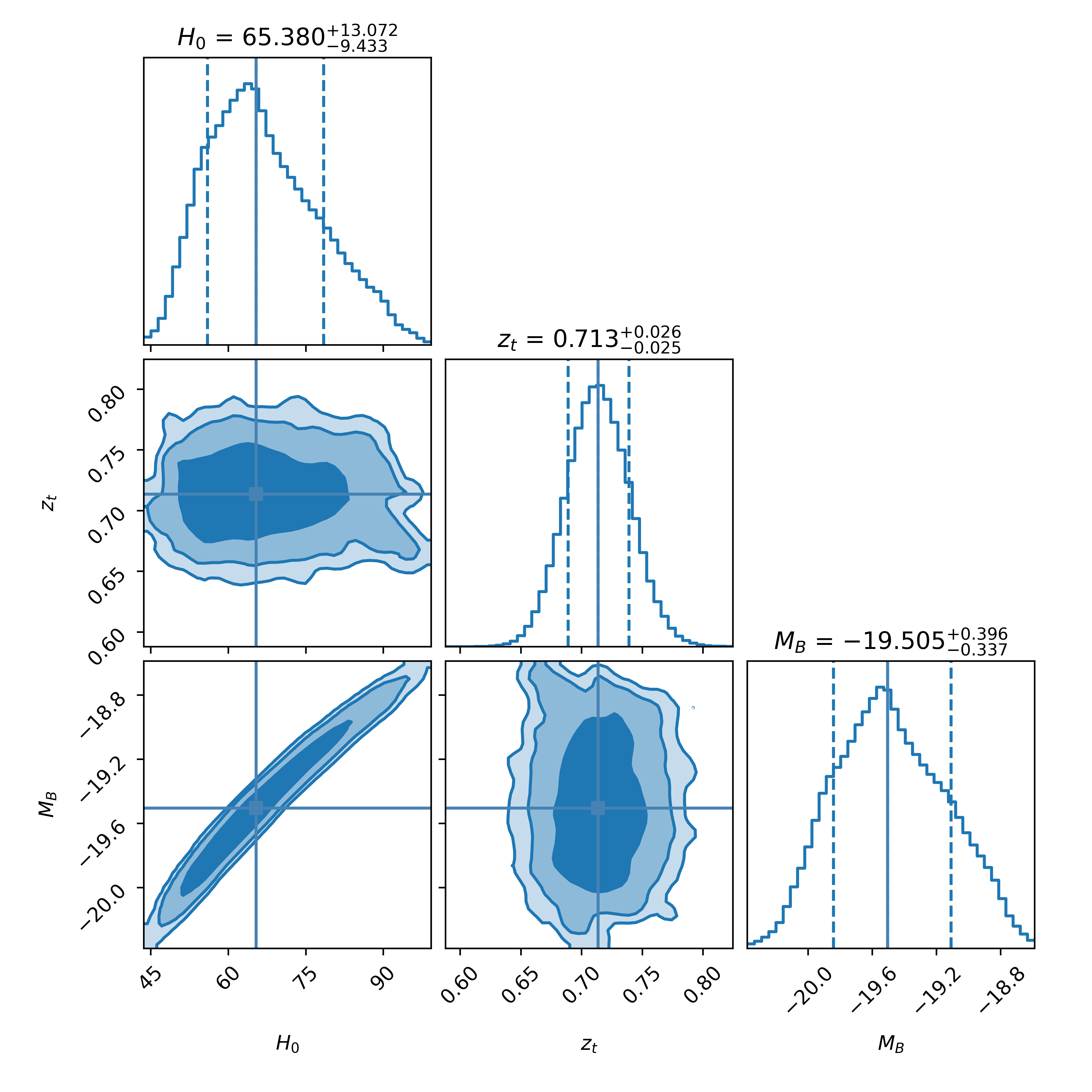}
\end{minipage}%
\hfill
\begin{minipage}{.475\textwidth}
  \centering
  \includegraphics[width=1.0\linewidth]{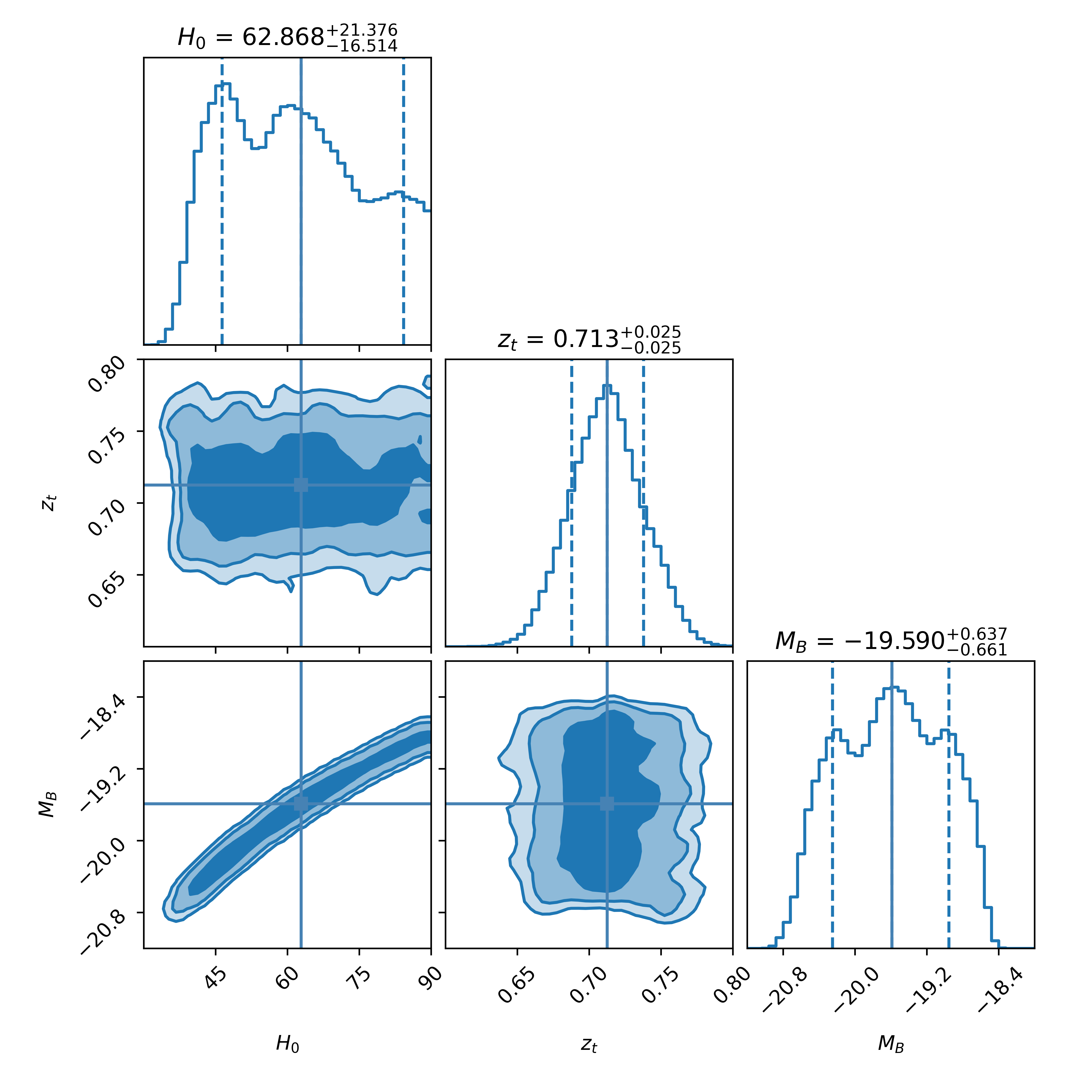}
\end{minipage}
\caption{The 1D and 2D posterior distributions of $H_0$, $z_{t}$ and $M_B$ obtained for $z_t$CDM model from GRBs+Type Ia SNe datasets with both constant $m~\&~c$ (\textbf{left figure)} and  dynamical $m~\&~c$ (\textbf{right figure}).}
    \label{csmly_w_GRB_SNIa_ztCDM_fig}
\end{figure}

The 1D and 2D posterior distributions of $H_0$, $z_{t}$ and $M_B$ with both constant $m~\&~c$ and  dynamical $m~\&~c$ are shown in Figure \ref{csmly_w_GRB_SNIa_ztCDM_fig}. 

Thus, we find that combining the datasets of GRBs and Type Ia SNe allows us to  put tighter  constraints on $z_t\sim 0.7$, which is  in concordance with the published value of $z_t$ obtained from different datasets \cite{2014JCAP...10..017M,2017ApJ...835...26F,2019ChPhC..43g5101L}. 
 
\section{Conclusions and Discussions}\label{conclusion}
As stated earlier, the main aim of this work is to calibrate the GRBs in a model-independent manner and then use it to put constraints on cosmological parameters. We calibrate the Amati relation which correlates the isotropic equivalent radiant energy $(E_{iso})$ and the spectral peak energy in the GRB rest frame $(E_p)$. The circularity issue that arises due to the lack of GRBs at low redshift has motivated  several authors to come up with model-independent approaches to calibrate this relation. In order to alleviate this problem, it has been proposed that the Type Ia SNe data available in redshift range of GRBs can be used to calibrate luminosity correlations of GRBs \cite{2008ApJ...685..354L,2010JCAP...03..005V}. It is assumed that the objects at the same redshift should have the same luminosity distance in any cosmology; hence the luminosity distance of the Type Ia SNe can be assigned to the GRBs existing at the same redshift or various model independent methods can  be used to obtain $d_L$ vs $\textit{z}$ plot. But the lack of Type Ia SNe beyond $z\sim 2$ makes it hard to calibrate the correlation at high redshift. One can only calibrate the relation for low redshift GRBs and extend it to high redshift. In the process of doing so, one has to make an assumption that the GRBs correlation is not evolving with redshift. As most of the GRBs are available at high redshifts, it becomes crucial to test this hypothesis. For this reason, cosmologists have been trying to test this relation with various data sets and methods.   \\

 L. Amati et al. proposed a novel technique to determine $d_L$ from the $H(z)$ data in a model independent way. They approximated the Hubble function corresponding to the OHD data points using a Bezier parametric curve obtained from a linear combination of Bernstein basis polynomials \cite{2019MNRAS.486L..46A}. This approach was also used in literature along with the enlarged Hubble data generated using various machine learning tools and BAO data to investigate the correlation \cite{2021MNRAS.503.4581L, 2023MNRAS.518.2247L}. In a recent work, a model independent approach has been used to constrain the Amati relation parameters and the intrinsic scatter \cite{2022JCAP...10..069G}. The authors used a non-parametric technique on the Galaxy cluster data to obtain $d_A$ corresponding to the GRBs redshift which is further used to determine $d_L$ assuming CDDR to be valid. Recently, N. Liang et al. (2022) calibrated the Amati relation of GRBs using Gaussian Process with the Type Ia SNe data. They obtain GRB Hubble diagram with the A219 and A118 samples of GRB and used it to test $\Lambda$CDM and $\omega$CDM models \cite{2022arXiv221102473N}. Z. Li et al also used GRB data to compare dark energy models \cite{2023MNRAS.521.4406L}. 
\vspace{3mm}\\ 
We divide this work into two parts as follows.

\begin{center}
    \textbf{Part I. Calibration of Gamma Ray Bursts}
\end{center}

 In this work, we use a non-parametric technique (Gaussian Process) to obtain the $d_L$ as a function of $z$ from the Hubble data having $32$ data points of Hubble parameter obtained from the differential age of the galaxies in the redshift range $0.0< z<2.0$. We use this luminosity distance and the bolometric fluence $S_{bolo}$, given in the GRB data, to calculate the bolometric isotropic equivalent radiant energy of GRBs ($E_{iso}$) upto $z\leq 2$. Further, the spectral peak energy ($E_p$) of the GRBs along with the $E_{\text{iso}}$ is used to put constraints on the Amati relation parameters and the intrinsic scatter. We believe that our work is advantageous over the earlier works for the following reasons:
 \begin{itemize}
\item We use $H(z)$ data obtained from the differential age of galaxies that does not rely on any cosmological assumption other than homogeneity and isotropy. In earlier work, a functional form of $H(z)$ (Bezier Polynomial) was used to obtain $d_L$ from $H(z)$ data. This method was applied to calibrate the Amati relation.
However in our work, we obtain $d_L$ using a non-parametric technique, namely Gaussian Process. Figure \ref{figure1} shows the variation of $d_L$ with redshift obtained on application of the Gaussian Process on the $H(z)$ data. This plot provides us the $d_L$ values at $z$ corresponding to the GRB data upto $z\sim 2$. The $1\sigma$, $2\sigma$ and $3\sigma$ contours of the Amati relation parameters and intrinsic scatter along with their best fit values are shown in Figure \ref{figure2}. We find that our results are in good concordance with the recent work in which authors have used galaxy cluster data to constrain the Amati relation parameters (See Table \ref{table1}). 

\item Recently N. Liang et al. used GP with Type Ia SNe data till redshift $z<1.4$ \cite{2022arXiv221102473N} and G. Govindaraj et al. with Galaxy cluster data upto $z< 0.9$ \cite{2022JCAP...10..069G}, to calibrate the Amati relation. In the latter, authors have to assume that the Cosmic Distance Duality Relation (CDDR) is valid to get the luminosity distance from the $d_A$ obtained from Galaxy cluster data. \textit{It is important to note that we do not make any such cosmological assumption in this work and hence making our analysis purely model independent}. Also $H(z)$ data is in redshift range $0 < z <2$ making it possible to test the validity of correlation upto comparatively higher redshift which is very crucial as most of the GRBs exist at high redshift. Another point to note is that by using this method, the errors in the parameters are smaller than the errors obtained by using the Galaxy Cluster data method as can be seen from Table \ref{table1}.
 
\item Further, in order to examine whether the Amati relation varies with cosmological redshift, we follow the technique of  \cite{2007MNRAS.379L..55L}. We divide the  sample of 200 GRBs out of 220 GRBs into five groups up to redshift $3.85$. To clarify, we arrange the GRBs according to their redshifts and distribute them into five bins based on a low-to-high range of redshift values with each bin  having 40 GRBs. 
We then  calibrate the Amati relation in each group and estimate the Amati relation parameters at the  mean redshift ${<}z{>}$ of given bin. For this, we use the data of old astrophysical objects and fit a  third order polynomial to estimate the luminosity distance. Once we obtain the  Amati relation parameters, i.e., $m$ and $c$ of each bin, we linear fit to $m-{<}z{>}$ and $c-{<}z{>}$. By analyzing the plot of $m-{<}z{>}$ and $c-{<}z{>}$, we find that $m$ and $c$  strongly evolve with the redshift which indicates the evolution of GRBs with the cosmological redshift.  L. Li (2007) also used the same methodology to study the redshift variation of Amati relation  with 48 GRBs only and concluded that GRBs evolve with the cosmological redshift \cite{2007MNRAS.379L..55L}.

\end{itemize} 
 
\begin{center}
    \textbf{Part II. Cosmological Parameters Estimation}
\end{center}
Due to their high luminosity,  GRBs have the potential to serve as alternate  probes to  Type Ia SNe for studying the evolution of the universe. After the observations of GRBs occurring at cosmological distances, various attempts  have been made to use GRBs correlations  to constrain cosmological models. In a recent study  \cite{2021JCAP...09..042K}, the Amati relation is employed to obtain constraints on both the correlation and cosmological model parameters. In their work, they focused on spatially-flat or non-flat $\Lambda$CDM, $X$CDM, $\phi$CDM models and put constraints on these models by using GRB data alone as well as the combined data of GRBs with BAO and $H(z)$. Similar work  has also been carried by \cite{2022arXiv221102473N}, where they calibrated the Amati relation using Type Ia SNe and investigate cosmology by adopting a joint analysis with GRBs and $H(z)$ datasets. \\

After calibrating GRBs from Part-I, in this part, we use the constant Amati relation as well as the  dynamical Amati relation and study  different cosmological models. In this work, we use three cosmological models: $\Lambda$CDM, $q(z)$ parametrization, and $z_t$CDM models. {\textit{Till now, only $\Lambda$CDM, $\phi$CDM and $X$CDM models have been studied with GRBs. In this work, we have extended the previous work by using different models, viz., $q(z)$ parametrization and $z_t$CDM models. In these two models, we study the deceleration parameter $(q(z))$ and transition redshift $(z_t)$ and put constraints on these parameters. 
Another notable point is that the constraints obtained earlier on cosmological parameters are carried out by using constant Amati relation only. In the present work, along with constant Amati relation, we test the impact of dynamical Amati relation on cosmological parameters of different cosmological models. }} Finally, we also test these cosmological models with a joint analysis of GRBs and Type Ia SNe datasets. 

A brief summary of the results is as follows:

\begin{itemize}
    \item \textbf{$\Lambda$CDM Model}: 
    \begin{itemize}
        \item {\textbf{GRB data:}}  The bounds on $H_0$ and $\Omega_{m0}$ are very weak as the error bars are quite large in both the cases, i.e, constant and dynamical m and c. It is observed that the dynamical m \& c supports a higher value of matter density.
        \item {\textbf{GRB+Type Ia SNe data:}} A joint analysis of the GRBs+Type Ia SNe provides more restrictive or tighter constraints on $\Omega_{m0}$ compared to the constraints from the GRBs data alone. Our analysis shows that the obtained value of $\Omega_{m0}$  is  consistent with the value obtained from the Type Ia SNe analysis \cite{2018ApJ...859..101S}. 
    \end{itemize}
     
    \item \textbf{$q(z)$ Parametrizations}: In this model, we consider three different parametrizations of the deceleration parameter.
    \begin{itemize}
        \item {\textbf{GRB data:}} 
      In all the parametrizations, it is observed that error bars on the model parameters ($q_i$ and $H_0$) are reduced when we work with the dynamical Amati relation. 
          
        \item {\textbf{GRB+Type Ia SNe data:}} Combining GRB data with Type Ia SNe data gives tighter constraints on all coefficients of the three parametrizations and we get a well constrained value for all model parameters (with both the constant and dynamical Amati parameters).
    \end{itemize}
    In all the parametrizations of $q(z)$, we find a negative value of $q_0$. This negative value suggests that the cosmic expansion is accelerating, which is the case in the current era of the universe. This is consistent with the observations of distant supernovae and the cosmic microwave background radiation, which indicate that the expansion of the universe is accelerating \cite{1998AJ....116.1009R, 1999ApJ...517..565P}.
     
    \item \textbf{$z_t$CDM Model:} In this model, we attempt to put constraints on the transition redshift, $z_t$. 
    \begin{itemize}
\item {\textbf{GRB data:}}  On comparing results for constant and dynamical m \& c relation, we find that $z_t$ decreases marginally. 
     
\item {\textbf{GRB+Type Ia SNe data:}} Combining GRB data with Type Ia SNe leads to a tighter constraint on $z_t\sim 0.7$, which is in concordance with the obtained value of $z_t$ from different datasets \cite{2014JCAP...10..017M,2017ApJ...835...26F,2019ChPhC..43g5101L}. Also the constrained value of $z_t$ remains same for both the cases (constant and dynamical m \& c).
\end{itemize}
    
\item If one were to conduct an analysis of GRBs+Type Ia SNe, the value of $H_0$, which represents the current expansion rate of the universe and ($M_B)$, the absolute magnitude of the supernova, would be free parameters to be determined. As expected in all the models, $H_0$ and $M_B$ show correlated behaviour. We also observe that there is very mild variation or no change in the  model parameters $(q_i, \Omega_{m0}$ \& $z_t$) values when we compare the  results obtained from constant m \& c with the dynamical relation. The error bars on model parameters reduce significantly when we work with GRB+Type Ia SNe.\\ 
\end{itemize}

The aforementioned evolutionary effect in GRB correlations is still under debate and several groups have been investigating the relation with different techniques and datasets \cite{2007MNRAS.379L..55L, 2008MNRAS.391..411B, 2008MNRAS.387..319G, 2009MNRAS.394L..31T, 2011MNRAS.415.3423W,2022MNRAS.516.2575J}. Lin et al. reported reasonable evidence of evolution with redshift for four relations and Wang et al. also found the same result for the Amati relation \cite{2016MNRAS.455.2131L, 2017ApJ...836..103W}. However some authors have claimed that there is no redshift evolution in the Amati relation \cite{2017A&A...598A.112D, 2021A&A...651L...8D, 2021JCAP...09..042K}. As a result, it becomes important to further examine the GRB relation in order to consider GRBs as standard candles. For that reason, the Space-based multiband astronomical Variable Objects Monitor (SVOM), a Sino-French mission which is expected to be launched in the middle of 2023 and is intended to study Gamma-Ray Bursts (GRBs) is of  great importance \cite{galaxies9040113}. Another space mission, Transient High Energy Sky and Early Universe Surveyor (THESEUS) is planned to be launched in 2032 \cite{2019eeu..confE...1A}. It aims to exploit the high redshift Gamma Ray Bursts in order to explore the early universe. In  light of these next generation missions, it is expected that we would have data that could provide better insight on the use of  GRBs as  standard candles.

\section*{Acknowledgements}  
Authors are thankful to the reviewer for very useful comments. D. Kumar is supported by an INSPIRE Fellowship under the reference number: IF180293 [SRF], DST India. N. Rani and D. Kumar acknowledge facilities provided by the IUCAA Centre for Astronomy Research and Development (ICARD), University of Delhi. Authors would also like to thank Akshay Rana for useful discussions. In this work some of the figures were created with \textbf{ {\texttt{corner}}} \cite{corner}, \textbf{ {\texttt{numpy}}}  \cite{numpy}  and \textbf{{\texttt{matplotlib}}} \cite{matplotlib} Python modules and to estimate parameters we used the publicly available MCMC algorithm  \textbf{ {\texttt{emcee}}} \cite{emcee}.  We are grateful to Sunny Vagnozzi for compiling the high-z OAO catalogue and sharing it with us.  
\bibliography{references_drsn}

\begin{thebibliography}{10}
\newcommand{\enquote}[1]{``#1''}

\bibitem{1998AJ....116.1009R}
A.~G. {Riess} \emph{et~al.}
\newblock {\emph{{Observational Evidence from Supernovae for an Accelerating
  Universe and a Cosmological Constant}}.}
\newblock \aj, \textbf{116}, 1009,  (1998).

\bibitem{1999ApJ...517..565P}
S.~{Perlmutter} \emph{et~al.}
\newblock {\emph{{Measurements of {\ensuremath{\Omega}} and
  {\ensuremath{\Lambda}} from 42 High-Redshift Supernovae}}.}
\newblock \apj, \textbf{517}, 565,  (1999).

\bibitem{2005ApJ...633..560E}
D.~J. {Eisenstein} \emph{et~al.}
\newblock {\emph{{Detection of the Baryon Acoustic Peak in the Large-Scale
  Correlation Function of SDSS Luminous Red Galaxies}}.}
\newblock \apj, \textbf{633}, 560,  (2005).

\bibitem{2004PhRvD..69j3501T}
M.~{Tegmark} \emph{et~al.}
\newblock {\emph{{Cosmological parameters from SDSS and WMAP}}.}
\newblock \prd, \textbf{69}, 103501,  (2004).

\bibitem{2015PhR...561....1K}
P.~{Kumar} and B.~{Zhang}.
\newblock {\emph{{The physics of gamma-ray bursts \& relativistic jets}}.}
\newblock Phys. Rep., \textbf{561}, 1,  (2015).

\bibitem{1993ApJ...413L.101K}
C.~{Kouveliotou}, C.~A. {Meegan}, G.~J. {Fishman}, N.~P. {Bhat}, M.~S.
  {Briggs}, T.~M. {Koshut}, W.~S. {Paciesas} and G.~N. {Pendleton}.
\newblock {\emph{{Identification of Two Classes of Gamma-Ray Bursts}}.}
\newblock \apjl, \textbf{413}, L101,  (1993).

\bibitem{2007PhR...442..166N}
E.~{Nakar}.
\newblock {\emph{{Short-hard gamma-ray bursts}}.}
\newblock Phys. Rept., \textbf{442}, 166,  (2007).

\bibitem{2006ARA&A..44..507W}
S.~E. {Woosley} and J.~S. {Bloom}.
\newblock {\emph{{The Supernova Gamma-Ray Burst Connection}}.}
\newblock \araa, \textbf{44}, 507,  (2006).

\bibitem{2002A&A...390...81A}
L.~{Amati} \emph{et~al.}
\newblock {\emph{{Intrinsic spectra and energetics of BeppoSAX Gamma-Ray Bursts
  with known redshifts}}.}
\newblock \aap, \textbf{390}, 81,  (2002).

\bibitem{2006MNRAS.372..233A}
L.~{Amati}.
\newblock {\emph{{The E$_{p,i}$-E$_{iso}$ correlation in gamma-ray bursts:
  updated observational status, re-analysis and main implications}}.}
\newblock \mnras, \textbf{372}, 233,  (2006).

\bibitem{2008MNRAS.391..577A}
L.~{Amati}, C.~{Guidorzi}, F.~{Frontera}, M.~{Della Valle}, F.~{Finelli},
  R.~{Landi} and E.~{Montanari}.
\newblock {\emph{{Measuring the cosmological parameters with the
  E$_{p,i}$-E$_{iso}$ correlation of gamma-ray bursts}}.}
\newblock \mnras, \textbf{391}, 577,  (2008).

\bibitem{2009A&A...508..173A}
L.~{Amati}, F.~{Frontera} and C.~{Guidorzi}.
\newblock {\emph{{Extremely energetic Fermi gamma-ray bursts obey spectral
  energy correlations}}.}
\newblock \aap, \textbf{508}, 173,  (2009).

\bibitem{2004ApJ...616..331G}
G.~{Ghirlanda}, G.~{Ghisellini} and D.~{Lazzati}.
\newblock {\emph{{The Collimation-corrected Gamma-Ray Burst Energies Correlate
  with the Peak Energy of Their {\ensuremath{\nu}}F$_{{\ensuremath{\nu}}}$
  Spectrum}}.}
\newblock \apj, \textbf{616}, 331,  (2004).

\bibitem{2004ApJ...609..935Y}
D.~{Yonetoku}, T.~{Murakami}, T.~{Nakamura}, R.~{Yamazaki}, A.~K. {Inoue} and
  K.~{Ioka}.
\newblock {\emph{{Gamma-Ray Burst Formation Rate Inferred from the Spectral
  Peak Energy-Peak Luminosity Relation}}.}
\newblock \apj, \textbf{609}, 935,  (2004).

\bibitem{2010A&A...511A..43G}
G.~{Ghirlanda}, L.~{Nava} and G.~{Ghisellini}.
\newblock {\emph{{Spectral-luminosity relation within individual Fermi gamma
  rays bursts}}.}
\newblock \aap, \textbf{511}, A43,  (2010).

\bibitem{2005ApJ...633..611L}
E.~{Liang} and B.~{Zhang}.
\newblock {\emph{{Model-independent Multivariable Gamma-Ray Burst Luminosity
  Indicator and Its Possible Cosmological Implications}}.}
\newblock \apj, \textbf{633}, 611,  (2005).

\bibitem{2022ApJ...924...97W}
F.~Y. {Wang}, J.~P. {Hu}, G.~Q. {Zhang} and Z.~G. {Dai}.
\newblock {\emph{{Standardized Long Gamma-Ray Bursts as a Cosmic Distance
  Indicator}}.}
\newblock \apj, \textbf{924}, 97,  (2022).

\bibitem{2019MNRAS.486L..46A}
L.~{Amati}, R.~{D'Agostino}, O.~{Luongo}, M.~{Muccino} and M.~{Tantalo}.
\newblock {\emph{{Addressing the circularity problem in the E$_{p}$-E$_{iso}$
  correlation of gamma-ray bursts}}.}
\newblock \mnras, \textbf{486}, L46,  (2019).

\bibitem{2020MNRAS.499..391K}
N.~{Khadka} and B.~{Ratra}.
\newblock {\emph{{Constraints on cosmological parameters from gamma-ray burst
  peak photon energy and bolometric fluence measurements and other data}}.}
\newblock \mnras, \textbf{499}, 391,  (2020).

\bibitem{2022ApJ...931...50L}
Y.~{Liu}, F.~{Chen}, N.~{Liang}, Z.~{Yuan}, H.~{Yu} and P.~{Wu}.
\newblock {\emph{{The Improved Amati Correlations from Gaussian Copula}}.}
\newblock \apj, \textbf{931}, 50,  (2022).

\bibitem{2008ApJ...685..354L}
N.~{Liang}, W.~K. {Xiao}, Y.~{Liu} and S.~N. {Zhang}.
\newblock {\emph{{A Cosmology-Independent Calibration of Gamma-Ray Burst
  Luminosity Relations and the Hubble Diagram}}.}
\newblock \apj, \textbf{685}, 354,  (2008).

\bibitem{2008MNRAS.391L...1K}
Y.~{Kodama}, D.~{Yonetoku}, T.~{Murakami}, S.~{Tanabe}, R.~{Tsutsui} and
  T.~{Nakamura}.
\newblock {\emph{{Gamma-ray bursts in $1.8 < z < 5.6$ suggest that the time
  variation of the dark energy is small}}.}
\newblock \mnras, \textbf{391}, L1,  (2008).

\bibitem{2010PASJ...62.1495Y}
D.~{Yonetoku}, T.~{Murakami}, R.~{Tsutsui}, T.~{Nakamura}, Y.~{Morihara} and
  K.~{Takahashi}.
\newblock {\emph{{Possible Origins of Dispersion of the Peak Energy-Brightness
  Correlations of Gamma-Ray Bursts}}.}
\newblock \pasj, \textbf{62}, 1495,  (2010).

\bibitem{2022arXiv221102473N}
N.~{Liang}, Z.~{Li}, X.~{Xie} and P.~{Wu}.
\newblock {\emph{{Calibrating Gamma-Ray Bursts by Using a Gaussian Process with
  Type Ia Supernovae}}.}
\newblock \apj, \textbf{941}, 84,  (2022).

\bibitem{2023MNRAS.518.2247L}
O.~{Luongo} and M.~{Muccino}.
\newblock {\emph{{Intermediate redshift calibration of gamma-ray bursts and
  cosmic constraints in non-flat cosmology}}.}
\newblock \mnras, \textbf{518}, 2247,  (2023).

\bibitem{2021MNRAS.501.3515M}
A.~{Montiel}, J.~I. {Cabrera} and J.~C. {Hidalgo}.
\newblock {\emph{{Improving sampling and calibration of gamma-ray bursts as
  distance indicators}}.}
\newblock \mnras, \textbf{501}, 3515,  (2021).

\bibitem{2022arXiv220813700M}
M.~{Muccino}, O.~{Luongo} and D.~{Jain}.
\newblock {\emph{{Constraints on the transition redshift from the calibrated
  Gamma-ray Burst $E_{\rm p}$-$E_{\rm iso}$ correlation}}.}
\newblock arXiv e-prints, arXiv:2208.13700,  (2022).

\bibitem{2021Galax...9...77L}
O.~{Luongo} and M.~{Muccino}.
\newblock {\emph{{A Roadmap to Gamma-Ray Bursts: New Developments and
  Applications to Cosmology}}.}
\newblock Galaxies, \textbf{9}, 77,  (2021).

\bibitem{2022JCAP...10..069G}
G.~{Govindaraj} and S.~{Desai}.
\newblock {\emph{{Low redshift calibration of the Amati relation using galaxy
  clusters}}.}
\newblock \jcap, \textbf{10}, 069,  (2022).

\bibitem{2010ApJ...714.1347S}
L.~{Samushia} and B.~{Ratra}.
\newblock {\emph{{Constraining Dark Energy with Gamma-ray Bursts}}.}
\newblock \apj, \textbf{714}, 1347,  (2010).

\bibitem{2015GReGr..47..141L}
J.~{Liu} and H.~{Wei}.
\newblock {\emph{{Cosmological models and gamma-ray bursts calibrated by using
  Pad{\'e} method}}.}
\newblock General Relativity and Gravitation, \textbf{47}, 141,  (2015).

\bibitem{2016MNRAS.455.2131L}
H.-N. {Lin}, X.~{Li} and Z.~{Chang}.
\newblock {\emph{{Model-independent distance calibration of high-redshift
  gamma-ray bursts and constrain on the {\ensuremath{\Lambda}}CDM model}}.}
\newblock \mnras, \textbf{455}, 2131,  (2016).

\bibitem{2022MNRAS.514.1828D}
M.~G. {Dainotti}, V.~{Nielson}, G.~{Sarracino}, E.~{Rinaldi}, S.~{Nagataki},
  S.~{Capozziello}, O.~Y. {Gnedin} and G.~{Bargiacchi}.
\newblock {\emph{{Optical and X-ray GRB Fundamental Planes as cosmological
  distance indicators}}.}
\newblock \mnras, \textbf{514}, 1828,  (2022).

\bibitem{2023MNRAS.518.2201D}
M.~G. {Dainotti}, A.~{\L}. {Lenart}, A.~{Chraya}, G.~{Sarracino},
  S.~{Nagataki}, N.~{Fraija}, S.~{Capozziello} and M.~{Bogdan}.
\newblock {\emph{{The gamma-ray bursts fundamental plane correlation as a
  cosmological tool}}.}
\newblock \mnras, \textbf{518}, 2201,  (2023).

\bibitem{2023arXiv230213887D}
S.-S. {Du}, J.-J. {Wei}, Z.-Q. {You}, Z.-C. {Chen}, Z.-H. {Zhu} and E.-W.
  {Liang}.
\newblock {\emph{{Model-Independent Determination of $H_0$ and $\Omega_{K,0}$
  using Time-Delay Galaxy Lenses and Gamma-Ray Bursts}}.}
\newblock arXiv e-prints, arXiv:2302.13887,  (2023).

\bibitem{2017A&A...598A.112D}
M.~{Demianski}, E.~{Piedipalumbo}, D.~{Sawant} and L.~{Amati}.
\newblock {\emph{{Cosmology with gamma-ray bursts. I. The Hubble diagram
  through the calibrated E$_{p,i}$-E$_{iso}$ correlation}}.}
\newblock \aap, \textbf{598}, A112,  (2017).

\bibitem{2019arXiv191108228D}
M.~{Demianski}, E.~{Piedipalumbo}, D.~{Sawant} and L.~{Amati}.
\newblock {\emph{{Prospects of high redshift constraints on dark energy models
  with the $E_p$- $E_{iso}$ correlation in long Gamma Ray Bursts}}.}
\newblock arXiv e-prints, arXiv:1911.08228,  (2019).

\bibitem{2021JCAP...09..042K}
N.~{Khadka}, O.~{Luongo}, M.~{Muccino} and B.~{Ratra}.
\newblock {\emph{{Do gamma-ray burst measurements provide a useful test of
  cosmological models?}}}
\newblock \jcap, \textbf{09}, 042,  (2021).

\bibitem{2022MNRAS.512..439C}
S.~{Cao}, M.~{Dainotti} and B.~{Ratra}.
\newblock {\emph{{Standardizing Platinum Dainotti-correlated gamma-ray bursts,
  and using them with standardized Amati-correlated gamma-ray bursts to
  constrain cosmological model parameters}}.}
\newblock \mnras, \textbf{512}, 439,  (2022).

\bibitem{2022MNRAS.510.2928C}
S.~{Cao}, N.~{Khadka} and B.~{Ratra}.
\newblock {\emph{{Standardizing Dainotti-correlated gamma-ray bursts, and using
  them with standardized Amati-correlated gamma-ray bursts to constrain
  cosmological model parameters}}.}
\newblock \mnras, \textbf{510}, 2928,  (2022).

\bibitem{2021MNRAS.507..730H}
J.~P. {Hu}, F.~Y. {Wang} and Z.~G. {Dai}.
\newblock {\emph{{Measuring cosmological parameters with a luminosity-time
  correlation of gamma-ray bursts}}.}
\newblock \mnras, \textbf{507}, 730,  (2021).

\bibitem{2022ApJ...935....7L}
Y.~{Liu}, N.~{Liang}, X.~{Xie}, Z.~{Yuan}, H.~{Yu} and P.~{Wu}.
\newblock {\emph{{Gamma-Ray Burst Constraints on Cosmological Models from the
  Improved Amati Correlation}}.}
\newblock \apj, \textbf{935}, 7,  (2022).

\bibitem{2016A&A...585A..68W}
J.~S. {Wang}, F.~Y. {Wang}, K.~S. {Cheng} and Z.~G. {Dai}.
\newblock {\emph{{Measuring dark energy with the E$_{iso}$ - E$_{p}$
  correlation of gamma-ray bursts using model-independent methods}}.}
\newblock \aap, \textbf{585}, A68,  (2016).

\bibitem{2019ApJ...887...13F}
F.~{Fana Dirirsa} \emph{et~al.}
\newblock {\emph{{Spectral Analysis of Fermi-LAT Gamma-Ray Bursts with Known
  Redshift and their Potential Use as Cosmological Standard Candles}}.}
\newblock \apj, \textbf{887}, 13,  (2019).

\bibitem{1994ApJS...95..107W}
G.~{Worthey}.
\newblock {\emph{{Comprehensive Stellar Population Models and the
  Disentanglement of Age and Metallicity Effects}}.}
\newblock \apjs, \textbf{95}, 107,  (1994).

\bibitem{2011MNRAS.412.2183Tf}
D.~{Thomas}, C.~{Maraston} and J.~{Johansson}.
\newblock {\emph{{Flux-calibrated stellar population models of Lick
  absorption-line indices with variable element abundance ratios}}.}
\newblock \mnras, \textbf{412}, 2183,  (2011).

\bibitem{2012JCAP...08..006M}
M.~{Moresco} \emph{et~al.}
\newblock {\emph{{Improved constraints on the expansion rate of the Universe up
  to z \raisebox{-0.5ex}\textasciitilde 1.1 from the spectroscopic evolution of
  cosmic chronometers}}.}
\newblock \jcap, \textbf{08}, 006,  (2012).

\bibitem{2003ApJ...593..622J}
R.~{Jimenez}, L.~{Verde}, T.~{Treu} and D.~{Stern}.
\newblock {\emph{{Constraints on the Equation of State of Dark Energy and the
  Hubble Constant from Stellar Ages and the Cosmic Microwave Background}}.}
\newblock \apj, \textbf{593}, 622,  (2003).

\bibitem{2018MNRAS.476.1036M}
J.~{Maga{\~n}a}, M.~H. {Amante}, M.~A. {Garcia-Aspeitia} and V.~{Motta}.
\newblock {\emph{{The Cardassian expansion revisited: constraints from updated
  Hubble parameter measurements and type Ia supernova data}}.}
\newblock \mnras, \textbf{476}, 1036,  (2018).

\bibitem{2022ApJ...928L...4B}
N.~{Borghi}, M.~{Moresco} and A.~{Cimatti}.
\newblock {\emph{{Toward a Better Understanding of Cosmic Chronometers: A New
  Measurement of H(z) at $z \sim 0.7$}}.}
\newblock \apjl, \textbf{928}, L4,  (2022).

\bibitem{2022ApJ...928..165W}
J.-J. {Wei} and F.~{Melia}.
\newblock {\emph{{Exploring the Hubble Tension and Spatial Curvature from the
  Ages of Old Astrophysical Objects}}.}
\newblock \apj, \textbf{928}, 165,  (2022).

\bibitem{2018ApJ...859..101S}
D.~M. {Scolnic} \emph{et~al.}
\newblock {\emph{{The Complete Light-curve Sample of Spectroscopically
  Confirmed SNe Ia from Pan-STARRS1 and Cosmological Constraints from the
  Combined Pantheon Sample}}.}
\newblock \apj, \textbf{859}, 101,  (2018).

\bibitem{2019A&A...625A..15T}
I.~{Tutusaus}, B.~{Lamine} and A.~{Blanchard}.
\newblock {\emph{{Model-independent cosmic acceleration and redshift-dependent
  intrinsic luminosity in type-Ia supernovae}}.}
\newblock \aap, \textbf{625}, A15,  (2019).

\bibitem{2022JCAP...01..053K}
D.~{Kumar}, A.~{Rana}, D.~{Jain}, S.~{Mahajan}, A.~{Mukherjee} and R.~F.~L.
  {Holanda}.
\newblock {\emph{{A non-parametric test of variability of Type Ia supernovae
  luminosity and CDDR}}.}
\newblock \jcap, \textbf{01}, 053,  (2022).

\bibitem{2020PhRvD.101j3517B}
G.~{Benevento}, W.~{Hu} and M.~{Raveri}.
\newblock {\emph{{Can late dark energy transitions raise the Hubble
  constant?}}}
\newblock \prd, \textbf{101}, 103517,  (2020).

\bibitem{2006gpml.book.....R}
C.~E. {Rasmussen} and C.~K.~I. {Williams}.
\newblock {\emph{{Gaussian Processes for Machine Learning}}.}
\newblock MIT Press,  (2006).

\bibitem{2021EPJC...81..892O}
E.~{{\'O} Colg{\'a}in} and M.~M. {Sheikh-Jabbari}.
\newblock {\emph{{Elucidating cosmological model dependence with H$_{0}$}}.}
\newblock European Physical Journal C, \textbf{81}, 892,  (2021).

\bibitem{turner2002type}
M.~S. Turner and A.~G. Riess.
\newblock {\emph{Do type Ia supernovae provide direct evidence for past
  deceleration of the universe?}}
\newblock The Astrophysical Journal, \textbf{569}, 18,  (2002).

\bibitem{2020A&A...641A...6P}
{Planck Collaboration} \emph{et~al.}
\newblock {\emph{{Planck 2018 results. VI. Cosmological parameters}}.}
\newblock \aap, \textbf{641}, A6,  (2020).

\bibitem{2002ApJ...569...18T}
M.~S. {Turner} and A.~G. {Riess}.
\newblock {\emph{{Do Type Ia Supernovae Provide Direct Evidence for Past
  Deceleration of the Universe?}}}
\newblock \apj, \textbf{569}, 18,  (2002).

\bibitem{Riess_2004}
A.~G. {Riess} \emph{et~al.}
\newblock {\emph{{Type Ia Supernova Discoveries at z > 1 from the Hubble Space
  Telescope: Evidence for Past Deceleration and Constraints on Dark Energy
  Evolution}}.}
\newblock \apj, \textbf{607}, 665,  (2004).

\bibitem{2008MPLA...23.1939X}
L.~{Xu} and H.~{Liu}.
\newblock {\emph{{Constraints to Deceleration Parameters by Recent Cosmic
  Observations}}.}
\newblock Modern Physics Letters A, \textbf{23}, 1939,  (2008).

\bibitem{2022arXiv221204751D}
D.~{Dahiya} and D.~{Jain}.
\newblock {\emph{{Revisiting the epoch of cosmic acceleration}}.}
\newblock arXiv e-prints, arXiv:2212.04751,  (2022).

\bibitem{Velasquez-Toribio:2020had}
A.~M. Velasquez-Toribio and A.~d.~R. Magnago.
\newblock {\emph{{Observational constraints on the non-flat $\Lambda CDM$ model
  and a null test using the transition redshift}}.}
\newblock Eur. Phys. J. C, \textbf{80}, 562,  (2020).

\bibitem{emcee}
D.~{Foreman-Mackey}, D.~W. {Hogg}, D.~{Lang} and J.~{Goodman}.
\newblock {\emph{{emcee: The MCMC Hammer}}.}
\newblock \pasp, \textbf{125}, 306,  (2013).

\bibitem{corner}
D.~{Foreman-Mackey}.
\newblock {\emph{{corner.py: Scatterplot matrices in Python}}.}
\newblock The Journal of Open Source Software, \textbf{1}, 24,  (2016).

\bibitem{2007MNRAS.379L..55L}
L.-X. {Li}.
\newblock {\emph{{Variation of the Amati relation with cosmological redshift: a
  selection effect or an evolution effect?}}}
\newblock \mnras, \textbf{379}, L55,  (2007).

\bibitem{2014JCAP...10..017M}
J.~{Maga{\~n}a}, V.~H. {C{\'a}rdenas} and V.~{Motta}.
\newblock {\emph{{Cosmic slowing down of acceleration for several dark energy
  parametrizations}}.}
\newblock \jcap, \textbf{10}, 017,  (2014).

\bibitem{2017ApJ...835...26F}
O.~{Farooq}, F.~{Ranjeet Madiyar}, S.~{Crandall} and B.~{Ratra}.
\newblock {\emph{{Hubble Parameter Measurement Constraints on the Redshift of
  the Deceleration-Acceleration Transition, Dynamical Dark Energy, and Space
  Curvature}}.}
\newblock \apj, \textbf{835}, 26,  (2017).

\bibitem{2019ChPhC..43g5101L}
H.-N. {Lin}, X.~{Li} and L.~{Tang}.
\newblock {\emph{{Non-parametric reconstruction of dark energy and cosmic
  expansion from the Pantheon compilation of type Ia supernovae}}.}
\newblock Chinese Physics C, \textbf{43}, 075101,  (2019).

\bibitem{2010JCAP...03..005V}
V.~{Vitagliano}, J.-Q. {Xia}, S.~{Liberati} and M.~{Viel}.
\newblock {\emph{{High-redshift cosmography}}.}
\newblock \jcap, \textbf{03}, 005,  (2010).

\bibitem{2021MNRAS.503.4581L}
O.~{Luongo} and M.~{Muccino}.
\newblock {\emph{{Model-independent calibrations of gamma-ray bursts using
  machine learning}}.}
\newblock \mnras, \textbf{503}, 4581,  (2021).

\bibitem{2023MNRAS.521.4406L}
Z.~{Li}, B.~{Zhang} and N.~{Liang}.
\newblock {\emph{{Testing dark energy models with gamma-ray bursts calibrated
  from the observational H(z) data through a Gaussian process}}.}
\newblock \mnras, \textbf{521}, 4406,  (2023).

\bibitem{2008MNRAS.391..411B}
S.~{Basilakos} and L.~{Perivolaropoulos}.
\newblock {\emph{{Testing gamma-ray bursts as standard candles}}.}
\newblock \mnras, \textbf{391}, 411,  (2008).

\bibitem{2008MNRAS.387..319G}
G.~{Ghirlanda}, L.~{Nava}, G.~{Ghisellini}, C.~{Firmani} and J.~I. {Cabrera}.
\newblock {\emph{{The E$_{peak}$-E$_{iso}$ plane of long gamma-ray bursts and
  selection effects}}.}
\newblock \mnras, \textbf{387}, 319,  (2008).

\bibitem{2009MNRAS.394L..31T}
R.~{Tsutsui}, T.~{Nakamura}, D.~{Yonetoku}, T.~{Murakami}, S.~{Tanabe},
  Y.~{Kodama} and K.~{Takahashi}.
\newblock {\emph{{Constraints on w$_{0}$ and w$_{a}$ of dark energy from
  high-redshift gamma-ray bursts}}.}
\newblock \mnras, \textbf{394}, L31,  (2009).

\bibitem{2011MNRAS.415.3423W}
F.-Y. {Wang}, S.~{Qi} and Z.-G. {Dai}.
\newblock {\emph{{The updated luminosity correlations of gamma-ray bursts and
  cosmological implications}}.}
\newblock \mnras, \textbf{415}, 3423,  (2011).

\bibitem{2022MNRAS.516.2575J}
X.~D. {Jia}, J.~P. {Hu}, J.~{Yang}, B.~B. {Zhang} and F.~Y. {Wang}.
\newblock {\emph{{E $_{iso}$-E$_{p}$ correlation of gamma-ray bursts:
  calibration and cosmological applications}}.}
\newblock \mnras, \textbf{516}, 2575,  (2022).

\bibitem{2017ApJ...836..103W}
G.-J. {Wang}, H.~{Yu}, Z.-X. {Li}, J.-Q. {Xia} and Z.-H. {Zhu}.
\newblock {\emph{{Evolutions and Calibrations of Long Gamma-Ray-burst
  Luminosity Correlations Revisited}}.}
\newblock \apj, \textbf{836}, 103,  (2017).

\bibitem{2021A&A...651L...8D}
Y.~{Dai}, X.-G. {Zheng}, Z.-X. {Li}, H.~{Gao} and Z.-H. {Zhu}.
\newblock {\emph{{Redshift evolution of the Amati relation: Calibrated results
  from the Hubble diagram of quasars at high redshifts}}.}
\newblock \aap, \textbf{651}, L8,  (2021).

\bibitem{galaxies9040113}
M.~G. Bernardini, B.~Cordier and J.~Wei.
\newblock {\emph{The SVOM Mission}.}
\newblock Galaxies, \textbf{9}, 4,  (2021).

\bibitem{2019eeu..confE...1A}
L.~{Amati}.
\newblock {\emph{{The Transient High-Energy Sky and Early Universe Surveyor
  (THESEUS)}}.}
\newblock In \enquote{The Extragalactic Explosive Universe: the New Era of
  Transient Surveys and Data-Driven Discovery,} 1 (2019).

\bibitem{numpy}
T.~E. Oliphant.
\newblock {\emph{A guide to NumPy}.}
\newblock Trelgol Publishing USA, \textbf{1},  (2006).

\bibitem{matplotlib}
J.~D. Hunter.
\newblock {\emph{Matplotlib: A 2D graphics environment}.}
\newblock Comput. Sci. Eng., \textbf{9}, 90,  (2007).

\end{thebibliography}
\bibliographystyle{Darshan_Custom_ref} 
\end{document}